\documentclass[amsmath, amssymb, preprintnumbers, noshowpacs, showkeys,aps,superscriptaddress,onecolumn]{revtex4}

\usepackage{color} 
\usepackage{hyperref}
\usepackage{slashed}
\usepackage{graphicx}
\usepackage{amsmath}
\usepackage{latexsym}
\usepackage{epstopdf}
\usepackage{amsmath}
\usepackage{enumitem}
\usepackage{amssymb,amsmath}
\usepackage{multirow}
\usepackage{mathtools}
\usepackage{bbm}
\usepackage{bm}
\usepackage{amsfonts}
\usepackage{amssymb}
\usepackage{calligra}
\usepackage{calrsfs}
\usepackage{makecell}
\usepackage{hyperref}
\usepackage{braket}

\newcommand{\beq}{\begin{equation}}
\newcommand{\eeq}{\end{equation}}

\newcommand{\be}{\begin{equation}}
\newcommand{\ee}{\end{equation}}

\begin{document}

\title{Superfluids, Fluctuations and Disorder}

\author{Alberto Cappellaro}
\email{cappellaro@pd.infn.it}
\affiliation{Dipartimento di Fisica e Astronomia “Galileo Galilei” and CNISM,
	Universit\`a di Padova, via Marzolo 8, 35131 Padova, Italy}
\author{Luca Salasnich}
\affiliation{Dipartimento di Fisica e Astronomia “Galileo Galilei” and CNISM,
	Universit\`a di Padova, via Marzolo 8, 35131 Padova, Italy}
\affiliation{CNR-INO, via Nello Carrara, 1 - 50019 Sesto Fiorentino, Italy}
\date{\today{}}

\begin{abstract}
We present a field-theory description of ultracold bosonic atoms 
in presence of a disordered external potential. 
By means of functional integration techniques,
we aim to investigate and review the interplay between disordered 
energy landscapes and fluctuations, both thermal and quantum ones. 
Within the broken-symmetry phase,
up to the Gaussian level of approximation, the disorder 
contribution crucially modifies both the condensate depletion and the superfluid response.
Remarkably, it is found that the ordered (i.e. superfluid) phase can be destroyed 
also in regimes where the random external potential is suitable for a perturbative
analysis. 
We analyze the simplest case of quenched disorder and then we move to
present the implementation of the replica trick for ultracold bosonic systems. In both
cases, we discuss strengths and limitations of the reviewed approach, 
paying specific attention to possible
extensions and the most recent experimental outputs.
\end{abstract}


\maketitle

\section{Introduction}

Ultracold atomic gases have been the cornerstone of atomic physics since
1995 \cite{cornell-1995,ketterle-1995}, when Bose-Einstein condensation 
was experimentally achieved for the first time.
This groundbreaking result was the starting point of an ongoing
and intense research effort to improve atoms trapping and cooling procedures
\cite{pethick-book}.
Nowadays, modern laboratories have reached an exquisite control on the
relevant physical parameters for atomic gases, such as the number density 
or the strength and range of atom-atom interaction. At the same time, 
it is also possible to tune the coupling with the external environement,
enabling the observation of quantum dynamics without spurious smearing effects
\cite{langen-2015}. 

It is then natural to think of ultracold atomic gases as an 
extremely effective platform to probe quantum theories up to the
macroscopic scale. One of the most striking example is given by the Bose-Hubbard model
and its superfluid-insulator transition originally predicted in \cite{fisher-1989}.
This quantum phase transition was first observed in 2002  by loading
a repulsive condensate in an optical lattice confining potential \cite{greiner-2002}.
Similar achievements have been reached for low-dimensional quantum gases and their
remarkable phenomenology \cite{giamarchi-book,dalibard-review-2d}. Among a wide
and interesting literature, it is worth mentioning the first direct observation
of the Tonks-Girardeau regime in an optical lattice filled with alkali atoms
\cite{paredes-2004,kinoshita-2004} or the vortex proliferation associated to the
Berezinskii-Kosterlitz-Thouless physics in two-dimensional setups 
\cite{hadzibabic-2006,cornell-2007}. 
More recently, a growing interest has aroused on cold atoms as a promising
quantum simulation platform. Together with the refined experimental expertise, 
it has been shown that, within peculiar setups, the superfluid dynamics of atoms
obeys to equations which can be mapped to other ones appearing in high-energy physics or cosmology.
For instance, Rabi-coupled bosonic mixtures seem to be a burgeoning candidate to explore
inflationary dynamics and space-time expansion in table top experiments
\cite{marquardt-2015,brandt-2017,braden-2018}. Other proposals involve 
cold-atoms analog of quantum
gravity \cite{liberati-2006} or the implementation of system mimicking QCD features
\cite{sademelo-2018}. 

Restraining ourselves to the condensed matter domain, ultracold atoms
offer the interesting possibility to investigate the intriguing 
issue of quantum transport in a disordered environment
\cite{kamenev-review}. Indeed, from the dawn of
solid state theory it was clear that a well-founded investigation has to consider
the multifaceted role played by microscopic impurities and disordered external potential. 
For instance, common incandescent
lamp would not exist without a certain degree of impurity scattering. At the same time,
it is reasonable to expect a smearing of experimental data due to randomly wrinkled
substrate or the destruction of supertransport properties.
Examples of this \textit{parasitic} role can be traced back to dirty superconductors
or $^4$He diffusion in porous medium \cite{ma-1985,ma-1986}.
The complexity of this problem lies in the interplay between superfluidity
(i.e. the occuring of some kind of ordered phase) and the localization phenomenon first theorized
by Anderson in \cite{anderson-1959}. 

Given the broad interest and the cross-disciplinary relevance of this problem,
it is reasonable to turn towards cold atoms, since they offer reliable 
experimental protocols to engineer disordered external potential and control
the corresponding relevant parameters.
The \textit{dirty bosons} problem then gathers all the investigations addressing the 
properties of ultracold bosons subject to an external random potential. The interplay
between superfluidity and localization may lead to the observation of a new phase, the
so-called Bose glass \cite{fisher-1989}. 
Similarly to the Bose-Hubbard model in absence of disorder, optical lattices have been proved again 
to be a reliable setup to explore these exotic features \cite{damski-2003}.
While interfering counterpropagating beams create the first regular pattern,
randomness is added by superimposing a second optical lattice, providing a viable strategy
to control the degree of disorder \cite{schulte-2005,billy-2008,roati-2008}.
An alternative is given by the speckle optical potential, generated by the interference
of waves with different phases and amplitudes but with the same frequency
\cite{dainty-1980,goodman-book}. 
The resulting wave varies randomly in space and can be used to study the transport properties
of ultracold atoms in presence of static external disorder 
\cite{lye-2005,clement-2005,ghabour-2014}.

In order to tackle down the crucial aspects of the dirty bosons problem, 
the relevant theoretical tools can
be divided in two different subsets. At first, one can adopt the point of view
of Bogoliubov theory, treating disorder in a perturbative way, similarly to
quantum and thermal fluctuations in the pure system \cite{stoof-book,salasnich-2016}.
Since the first seminal work on this topic \cite{huang-1992,giorgini-1994,tauber-1997}, 
this approach has led us to a better understanding of the role played by disorder,
no matter how small, in condensed bosonic systems. The additional depletion to
condensate and superfluid was computed in \cite{huang-1992}, while in \cite{giorgini-1994}
the authors investigated the modifications induced by a random external potential to the sound
propagation. At the same time, critical properties are investigated by focusing, for instance,
on the condensation temperature shift driven by the interplay of disorder and interaction
\cite{lopatin-2002,falco-2007}. 

A non-perturbative approach paves the way to the exploration of the 
superfluid-glass (quantum) transition in bosonic ensembles. Moving from the
seminal renormalization-group analysis for one-dimensional systems 
\cite{giamarchi-1988,giamarchi-book}, different strategies and methods 
have been used to understand the phase diagram of dirty bosons in 
three and lower dimensions, such as the random-phase approximation \cite{navez-2007},
stochastic field-theory techniques and the replica formalism 
\cite{yukalov-2007,falco-2009,khellil-2016,khellil-2016-2}.

The ambition of this review is obviously more limited, since it is not possible
face in the main text all the topics we have mentioned during this introduction. 
Nevertheless, we aim to present, in a clear and pedagogical way, 
the crucial ingredients of a field-theory analysis
of superfluid bosons. Throughtout the proceeding of the paper, we are going to adopt the
functional integration framework, which is flexible enough to enable the implementation
of different approximation schemes \cite{altland-book}. Our discussion is organized in the following
way: in Sec. \ref{sec:2} we specify how one can build up a quantum field theory for superfluid bosons 
subject to a random confinement. In particular, one needs to clarify the way in which
disorder is introduced and characterized. 
In Sec. \ref{sec:3} we compute, up to the Gaussian level of approximation, 
the additional contribution to the condensate and the superfluid
component due to the presence of a quenched disordered potential. 
Moving to Sec. \ref{sec:4}, we present the core ideas at the foundation of the replica formalism, 
which allows to generalize our analysis to more complex disorder realization and does not
rely, in principle, upon the assumption of a perturbatively small disorder. Within
this scheme, we show how to compute the meaningful correlation functions and describe the superfluid
response of the system in Sec. \ref{sec:5}. Further comments and future perspective are addressed
in the Conclusions.

\section{Disorder in Field Theories}
\label{sec:2}

Let us begin by considering $N$ interacting bosonic particles of mass $m$ enclosed in
a $L^d$ volume and subject to an external \textit{disordered} potential 
$U_{\text{D}}(\mathbf{r})$. Thus, the Hamiltonian is simply given by
\begin{equation}
\hat{H} = \sum_{i=1}^{N} \bigg[-\frac{\hbar^2\nabla_{i}^2}{2m} 
+ \hat{U}_{\text{D}}(\mathbf{r}_i) \bigg]
+ \frac{1}{2}\sum_{i\neq j}\hat{V}\big(|\mathbf{r}_i-\mathbf{r}_j|\big)\;,
\label{hamiltoniana di partenza}
\end{equation}
with $\hat{V}\big(|\mathbf{r}_i-\mathbf{r}_j|\big)$ the two-body central interaction
potential.
The thermodynamic description of the system can be derived from the partition function.
Within the canonical ensemble, it is defined as
\begin{equation}
Z_N = \sum_{\lbrace n\rbrace} \exp\big( -\beta E_n\big)
\label{funzione di partizione canonica}
\end{equation}
where $\beta \equiv (k_B T)^{-1}$ and $\lbrace n\rbrace$
denoting the bosonic states (i.e. totally symmetric with respect to
particle exchange) 
with eigenenergy $E_n$ of the operator
$\hat{H}$ in Eq. \eqref{hamiltoniana di partenza}.
In order to adopt the functional integration formalism \cite{altland-book}, 
Eq. \eqref{hamiltoniana di partenza}
has to be expressed in second quantization. By adding $-\mu \hat{N}$, with $\mu$ the 
chemical potential and $\hat{N}$ the number operator, we can also move to the grand canonical 
ensemble. Here, a basis of coherent bosonic states can be chosen, leading to the imaginary-time
representation of the (grand canonical) partition function
\begin{equation}
\mathcal{Z} = \int\mathcal{D}[\psi,\psi^*] \,e^{-S[\psi,\psi^*]/\hbar}\;.
\label{funzione di partizione grancanonica}
\end{equation}
In the equation above, the Bose statistics is encoded in the periodic boundary condition for
the fields, namely $\psi(0) = \psi(\beta\hbar)$ and $\psi^*(0)=\psi^*(\beta\hbar)$. 
The Eucliden action appearing in Eq. \eqref{funzione di partizione grancanonica} is given by
\begin{equation}
\begin{aligned}
S[\psi,\psi^*] & = \int_0^{\beta\hbar}d\tau \int d^d\mathbf{r}\;
\bigg\lbrace 
\psi^*(\mathbf{r},\tau)\bigg[ \hbar\partial_{\tau} - \frac{\hbar^2\nabla^2}{2m} 
-\mu - U_{\text{D}}(\mathbf{r},\tau)\bigg]\psi (\mathbf{r},\tau)\\
&\qquad \qquad \qquad \qquad\qquad +\frac{1}{2}\int d^d\mathbf{r}'|\psi(\mathbf{r},\tau)|^2
V(|\mathbf{r}-\mathbf{r}'|) |\psi(\mathbf{r}',\tau)|^2
\bigg\rbrace\;.
\end{aligned}
\label{azione euclidea di partenza}
\end{equation}
We have now to clarify what is meant by a (quantum) field theory for
the disordered Bose gas. Also in the simplest case of non-interacting bosons, 
given a certain realization of the disordered confinement,
an exact diagonalization of Eq. \eqref{hamiltoniana di partenza} does not appear
as a feasible task.

A statistical approach is certainly more reasonable, since it allows, 
in principle, to characterize the system through a limited range of variables as, for instance,
the strength of the impurity potential leading to the disorder, or the scale 
of space-time potential fluctuations. This obviously implies that it is possible to average
(in a way to define) over different microscopic realization of the disorder potential.

This outline may seem very vague: how can we implement a theoretical description
in statistical terms?  Apparently, it is simple to imagine a set of $N$ impurities at
$\lbrace \mathbf{r}_i\rbrace_{i=1,\ldots , N}$ positions. When they are assumed equivalent, the 
resulting \textit{disorder} (i.e. impurity) potential takes the obvious form
$U_{\text{D}}(\mathbf{r}) = \sum_i U(\mathbf{r}-\mathbf{r}_i)$. 
Within this recipe 
the disorder average is equivalent to the integration over the whole 
configuration space of impurities, namely
\begin{equation}
\braket{\,\ldots\,}_{\text{dis}}\equiv 
L^{-{Nd}}\prod_{i=1}\int d^d\mathbf{r}_i \big (\ldots\big)\;,
\label{disorder average old style}
\end{equation}
a technical task not easy to implement in 
a functional formalism. 

As pointed out throughtout a vast literature 
(\cite{hertz-1985,nelson-1990,tauber-1997,schakel-1997,lopatin-2002,falco-2007,
lubensky-1975,grinstein-1976}, 
just to mention a few),
a more fruitful approach relies upon the statistical features of the disorder potential or,
in other words, a description in terms of 
probability density function (PDF) $\mathsf{P}(U_{\text{D}})$. 
It follows then that
\begin{equation}
\braket{\,\ldots\,}_{\text{dis}} \equiv \int \mathcal{D}[U_{\text{D}}] \mathsf{P}[U_{\text{D}}]
(\ldots)\;.
\label{disorder average con pdf}
\end{equation}
For instance, we can surely imagine a Gaussian distributed 
disorder, i.e.
\begin{equation}
\mathsf{P}[U_{\text{D}}] = \exp\bigg\lbrace
- \frac{1}{2\gamma^2} \int d^{d+1}\mathbf{s}\int d^{d+1}\mathbf{s}' \;U_{\text{D}}(\mathbf{s})
K^{-1}(\mathbf{s}-\mathbf{s}')U_{\text{D}}(\mathbf{s}')
\bigg\rbrace
\label{pdf gaussiana}
\end{equation}
where $\mathbf{s}=(\mathbf{r},\tau)$
and $d^{d+1}{\bf s} = d\tau \,d^{d}{\bf r}$. The expectation value of a central Gaussian distributed
variable is zero, 
i.e.
\begin{equation}
\langle U_D({\bf s}) \rangle_{\text{dis}} = 0 \; ,
\end{equation}
while its second (central) momentum corresponds to
\begin{equation}
\braket{U_{\text{D}}(\mathbf{s}) U_{\text{D}}(\mathbf{s}')}_{\text{dis}}=
\gamma^2\,K(\mathbf{s}-\mathbf{s}')\;.
\label{secondo momento centrale}
\end{equation}
For a $\delta$-correlated disorder
, where $K({\bf s} - {\bf s}') = \delta^{(d+1)}({\bf s} - {\bf s}')$ 
with $\delta^{(d+1)}({\bf s})$ the Dirac delta function in $d+1$ dimensions,
the above equation is further simplified in
\begin{equation}
\mathsf{P}[U_{\text{D}}] = 
\exp\bigg\lbrace
- \frac{1}{2\gamma^2}\int_{0}^{\beta\hbar}d\tau\int d^d\mathbf{r}\;
U^2_{\text{D}}(\mathbf{r},\tau)
\bigg\rbrace\;.
\label{pdf per disordine delta correlato}
\end{equation}
If we deal with atomic bosons in the superfluid phase
the atomic wavelength is 
much larger than the scattering range of impurities generating the disorder potential
$U_{\text{D}}(\mathbf{s})$. This physical situation clarifies the simplifying
assumption of a $\delta$-correlated disorder as in $\eqref{pdf per disordine delta correlato}$.
Moreover, in this section we explicitly consider the case of a quenched disorder,
where the time scale of $U_{\text{D}}$ is much longer than the thermodynamic one.
From a technical point of view this implies that $U_{\text{D}}$ retains only 
a spatial dependence on $\mathbf{r}$ and that the disorder average in 
Eq. \eqref{disorder average con pdf} has to be performed after the thermal one
\cite{falco-2007}. Thus, the average over a quenched disorder has to be intended as
\begin{equation}
\braket{\mathcal{O}[\psi,\psi^*]}_{\text{dis+th}} =
\int \mathcal{D}[U_{\text{D}}] \mathsf{P}[U_{\text{D}}]
\braket{\mathcal{O}[\psi,\psi^*]}_{\text{th},U_{\text{D}}}
\label{media quenched con pdf}
\end{equation}
where
\begin{equation}
\braket{\mathcal{O}[\psi,\psi^*]}_{\text{th},U_{\text{D}}}
=
\int\mathcal{D}[\psi,\psi^*]\frac{1}{\mathcal{Z}(U_{\text{D}})}
\mathcal{O}[\psi,\psi^*] e^{-S[\psi,\psi^*]/\hbar}
\label{media termica quenched}
\end{equation}
is the usual thermal average over the grand canonical ensemble for a 
given impurity configuration. Within the scheme outlined by
Eqs. \eqref{media quenched con pdf} and \eqref{media termica quenched},
$U_{\text{D}}(\mathbf{r})$ and $\psi(\mathbf{r},\tau)$ are not
being treated on the same footing, since the random field $U_{\text{D}}$
is fixed when we average over the bosonic field $\psi$. More precisely, 
Eq. \eqref{media termica quenched} is telling us to compute the partition function
over a precise microscopic realization of the system (i.e. with a given $U_{\text{D}}$).
Then, according to Eq. \eqref{media quenched con pdf}, we have to average all the
partition functions obtained in this way over the disorder field.

The quenched average presented above is certainly a viable scheme to adopt when the
disorder can be treated in a perturbative way. However, this is not always the case. 
It can occur a physical situation where one has to perform the disorder average early in
the calculation and/or the random field $U_{\text{D}}$ has to be considered at the same
level of $\psi$. In order to highlight the eventual technical obstacle, we make use of 
the source method to compute expectation values of physical observable. According to this
field-theory technique, the expectation value of a given observable $\mathcal{O}$
can be extracted by means of the source method
\cite{altland-book,schakel-book}.
Here, the source is intended as a term linear in the
fields and proportional to $J(\mathbf{r},\tau)$, namely
\begin{equation*}
\mathcal{Z}[J,J^*]= \int \mathcal{D}[\psi,\psi^*]
\;.\exp\bigg\lbrace
-\hbar^{-1}S[\psi,\psi^*] + \hbar^{-1}\int_0^{\beta\hbar}d\tau\int d^d\mathbf{r}
\big[J^*(\mathbf{r},\tau)\psi(\mathbf{r},\tau) + J(\mathbf{r},\tau)\psi^*(\mathbf{r},\tau) \big]
\bigg\rbrace\;.
\end{equation*}
A simple differentiation leads us to 
the pursued expectation value. The most immediate example is the chemical potential,
	indeed $\braket{N} = \beta^{-1}\partial_{\mu}\log\mathcal{Z}$.
The thermal average of $\mathcal{O}$ is defined as
\begin{equation}
\braket{\mathcal{O}}_{\text{th}} = - \frac{\delta}{\delta J}\log\mathcal{Z}\bigg|_{J=0}\;.
\label{media termica integrazione funzionale}
\end{equation}
As a consequence of Eq. \eqref{disorder average con pdf}, the joint thermal and disorder 
average reads
\begin{equation}
\braket{\mathcal{O}}_{\text{dis+th}}  = - \frac{\delta}{\delta J}
\braket{\log  \mathcal{Z}}_{\text{dis}}\bigg|_{J=0} \\
= - \int\mathcal{D}[U_{\text{D}}]\;
\frac{\mathsf{P}(U_{\text{D}})}{\mathcal{Z}[J=0,U_{\text{D}}]}
\frac{\delta}{\delta J} \mathcal{Z}[J,U_{\text{D}}]\bigg|_{J=0}\;.
\label{media con il metodo delle sorgenti}
\end{equation}
Since one has both to differentiate with respect to $J$ and integrate over $U_{\text{D}}$,
the disorder field appears both at the numerator and denominator.
Consequently, Eq. \eqref{media con il metodo delle sorgenti} can be challenging.

In the following sections we approach both the situations: first we consider the
case of a quenched disorder potential, then we move to a more general framework by
presenting the implementation of the replica trick for superfluid atomic bosons. 


\section{Perturbative approach to quenched disorder}

\label{sec:3}

\subsection{Thermodynamic picture and disorder-driven condensate depletion}

Let us consider the Euclidean action in Eq. \eqref{azione euclidea di partenza}
and apply the saddle-point method by stationarizing it, i.e.
\begin{equation}
\bigg[
\hbar\frac{\partial}{\partial \tau} - \frac{\hbar^2\nabla^2}{2m}-
(\mu+U_{\text{D}}(\mathbf{r},\tau)) + \int d^d\mathbf{r}'\,
|\psi(\mathbf{r}',\tau)|^2V(|\mathbf{r}-\mathbf{r}'|)
\bigg]\psi(\mathbf{r},\tau) = 0\;.
\label{equazione di schroedinger non-linear}
\end{equation}
In the case of a two-body contact interaction, we have 
\begin{equation}
V = g\;\delta^{(d)}(\mathbf{r})
\label{potenziale di contatto}
\end{equation}
where
$g = \tilde{V}(\mathbf{q}=0)$ and $\tilde{V}(\mathbf{q})$ is the Fourier transform of the
two-body potential. For a uniform configuration of the field, with
\begin{equation}
\psi(\mathbf{r},\tau) = \psi_0
\end{equation}
not depending on $\mathbf{r}$ and $\tau$, we get
\begin{equation}
\big[-(\mu + \braket{U_{\text{D}}}_{\text{dis}}) + g\psi_0^2\big]\psi_0
= 0 \Longrightarrow 
\begin{cases}
\psi_0^2= 0 \qquad \text{for $\mu + \braket{U_{\text{D}}}_{\text{dis}} \leq 0$} \\
\\
\psi_0^2 = (\mu + \braket{U_{\text{D}}}_{\text{dis}})/g 
\qquad \text{for $\mu + \braket{U_{\text{D}}}_{\text{dis}} > 0$}\;.
\end{cases}
\label{saddle point}
\end{equation}
This is approximation must not be underrated: actually, in presence of an external
potential $\psi_0$ should not be uniform. However, at low temperatures it seems reasonable
to assume that the wavelengths of atoms are much larger than the spatial variations
caused by the impurity potential.
	
By comparing the equations above with the ones for a pure system 
\cite{salasnich-2016}, 
it is immediate
to realize that, in principle, the disorder may shift the critical point of the superfluid
transition. However, if $U_{\text{D}}$ follows a centered Gaussian distribution like the one
in Eq. \eqref{pdf gaussiana}, this shift equates to zero. From the saddle-point result it
descends that the total density $n =  \psi^2_0$.
By identifying $\psi_0^2$ with the condensate density $n_0$, this
signals that, up to this (very rough) level of approximation, 
all the particles take part in the condensate within the 
broken-symmetry phase. 

In this phase, where a $\mathsf{U}(1)$
symmetry is spontaneously broken, fluctuations can be taken into account
by considering the following splitting of the field
\begin{equation}
\psi(\mathbf{r},\tau) = \psi_0 + \eta(\mathbf{r},\tau)
\label{splitting del campo 1}\;,
\end{equation}
with $\eta(\mathbf{r},\tau) \in \mathbb{C}$. 
Replacing the equation above in the Euclidean action of
Eq. \eqref{azione euclidea di partenza}, the disorder contribution reads
\begin{equation}
\mathcal{L}_{\text{dis}}[\psi,\psi^*,U_{\text{D}}] =
- U_{\text{D}}(\mathbf{r})\big|\psi(\mathbf{r},\tau)\big|^2 =
-U_{\text{D}}(\mathbf{r}) \big[ 
\psi_0^2 + |\eta(\mathbf{r},\tau)|^2 + v\eta(\mathbf{r},\tau) + v\eta^*(\mathbf{r},\tau)
\big]
\label{lagrangiana di disordine con splitting}
\end{equation}
where $\mathcal{L}$ denotes the (Euclidean) Lagrangian density, i.e. 
$S[\psi] = \int d\tau \int d^d\mathbf{r}\;\mathcal{L}[\psi]$. The first two terms 
give rise to an unobservable shift of the chemical potential, so we will neglect them.
Moreover, let us remark that we
are considering fluctuations 
over the saddle-point non-trivial 
configuration $\psi_0^2 = \mu/g$ with $\braket{U_{\text{D}}}_{\text{dis}} = 0$, but this is
not the \text{true} stationary point. Consequently, we are not allowed to neglect 
the linear terms in Eq. \eqref{lagrangiana di disordine con splitting}, not even up to
the Gaussian order in the fluctuation fields.
By introducing the column vectors
\begin{equation}
\Upsilon = \psi_0
\begin{pmatrix}
1 \\
1
\end{pmatrix}
\qquad \text{and}
\qquad
\chi=
\begin{pmatrix}
\eta \\
\eta^*
\end{pmatrix}
\label{vettori colonna sezione quenched}
\end{equation}
Eq. \eqref{lagrangiana di disordine con splitting} is brought to its more compact form
\begin{equation}
\mathcal{L}_{\text{dis}}[\eta,\eta^*,U_{\text{D}}] = -U_{\text{D}}(\mathbf{r})
\Upsilon^T \chi\;.
\label{lagrangiana di disordine versione compatta}
\end{equation}
At this point, we can replace the field $\psi(\mathbf{r},\tau)$ with 
Eq. \eqref{splitting del campo 1}. By retaining terms up to the quadratic order 
in $\eta$ and $\eta^*$, the grand canonical partition function reads
\begin{equation}
\begin{aligned}
\mathcal{Z}(U_{\text{D}}) & = e^{-S_{\text{mf}}[\psi_0,\mu]/\hbar}
\int \mathcal{D}[\chi] \exp\bigg\lbrace
-\frac{1}{\hbar}\int_0^{\beta\hbar}d\tau\int d^d\mathbf{r}\,\big[
\mathcal{L}_g^{\text{(pure)}} +\mathcal{L}_{\text{dis}}
\big]
\bigg\rbrace \\
&= e^{\beta L^d\mu^2/(2g)} \int \mathcal{D}[\chi]\exp
\bigg\lbrace 
- \frac{1}{\hbar}\int d^{d+1}\mathbf{s}\bigg[
\frac{1}{2}\chi^{\dagger}\mathcal{G}^{-1}(\partial_{\tau},\nabla)\chi - 
U_{\text{D}}(\mathbf{r})\Upsilon^T\chi
\bigg]
\bigg\rbrace\;,
\end{aligned}
\label{funzione di partizione gaussiana con disordine}
\end{equation}
with $\mathbf{s} = (\mathbf{r},\tau)$.
In the equation above, the integral in $\chi$ is Gaussian and, consequently, can be
computed exactly, leading us to the following result
\begin{equation}
\mathcal{Z}(U_{\text{D}}) = e^{-\beta \Omega_{\text{mf}}}
\; \big(\det \mathcal{G}^{-1}\big)^{-1} \exp\bigg\lbrace
\frac{1}{2}\int d^{d+1}\mathbf{s}\int d^{d+1}\mathbf{s}' U_{\text{D}}(\mathbf{r})
\Upsilon^T \mathcal{G}(\mathbf{s}-\mathbf{s}')\Upsilon U_{\text{D}}(\mathbf{r}')
\bigg\rbrace\;.
\label{funzione di partizione gaussiana fluttuazioni integrate via}
\end{equation}
Let us also recall that, from Eq. \eqref{saddle point} it is easy to
realize that the mean-field grand potential is simply
$\Omega_{\text{mf}}/L^d = -\mu^2/2g$.
The remaining steps of the calculation are easier in 
the Fourier space, where the (inverse) Gaussian propagator is the
following $2\times 2$ matrix
\begin{equation}
\mathcal{G}^{-1}(\mathbf{q},\omega)  = 
\begin{pmatrix}
-i\hbar\omega_n + \epsilon_q -\mu + 2g\psi_0^2 & g\psi_0^2 \\
& \\
g\psi_0^2 & +i\hbar\omega_n + \epsilon_q -\mu + 2g\psi_0^2
\end{pmatrix}
\label{inverso del porpagatore gaussiano spazio fourier alla schakel}
\end{equation}
with $\omega_{n} = 2\pi n/\beta$ the bosonic Matsubara frequencies and
$\epsilon_q = \hbar q^2/(2m)$.
 The (functional) determinant in 
Eq. \eqref{funzione di partizione gaussiana fluttuazioni integrate via}
can be handled more easily thanks to the identities
\begin{equation}
\big(\det \mathcal{G}^{-1}\big)^{-1} = \exp\big(-\text{tr}\log\mathcal{G}^{-1}\big)
= \exp\big(-\log\det\mathcal{G}^{-1}\big)\;.
\label{identita per il determinante funzionale}
\end{equation} 
By putting all these pieces together, 
Eq. \eqref{funzione di partizione gaussiana fluttuazioni integrate via}
is built by three blocks, namely
\begin{equation}
\mathcal{Z}(U_{\text{D}}) = e^{-\beta\Omega_{\text{mf}}} 
\exp\bigg\lbrace
\underbrace{-\frac{1}{2}\sum_{\mathbf{q}}\sum_n \log\det\mathcal{G}^{-1}(\mathbf{q},\omega_n)
	\bigg\rbrace}_{\text{Pure Gaussian term}}
\exp\bigg\lbrace
\underbrace{\frac{1}{2}\sum_{\mathbf{q}} \big|U_{\text{D}}(\mathbf{q})\big|^2 \Upsilon^T
	\mathcal{G}(\mathbf{q},0)\Upsilon}_{\text{disorder Gaussian term }}
\bigg\rbrace\;.
\label{funzione di partizione gaussiana fattorizzata fourier}
\end{equation}
The disorder contribution in the equation above does not depend on the
frequency, since the time scales of the quenched disordered are frozen, implying 
that the corresponding frequencies goes to zero. 
Concerning the pure Gaussian contribution, we have \cite{salasnich-2016}
\begin{equation}
\frac{1}{2\beta}\sum_n \log\det\mathcal{G}^{-1}(\mathbf{q},\omega_n)= 
\frac{1}{2\beta} \sum_n \log\big[ \beta^2\big(\hbar^2\omega_n^2 +E_q^2\big)\big]
= \frac{1}{2}\,E_q + \frac{1}{\beta}\log\big(1-e^{-\beta E_q}\big)\;,
\end{equation}
where a crucial role is played by the excitation spectrum of collective
excitations above the uniform saddle-point configuration, i.e.
\begin{equation}
E_q(\mu,\psi_0) = \sqrt{
	\big( 
	\epsilon_q -\mu + 2g\psi_0^2
	\big)^2 - g^2\psi_0^4
}\;.
\label{spettro delle fluttuazione sopra il punto sella}
\end{equation}
Focusing on the zero-temperature case, we find
\begin{equation}
\mathcal{Z}(U_{\text{D}}) = e^{-\beta(\Omega_{\text{mf}}+\Omega_g^{(0)})}
\; \exp\bigg\lbrace
\frac{1}{2}\sum_{\mathbf{q}} \big|U_{\text{D}}(\mathbf{q})\big|^2 \Upsilon^T
\mathcal{G}(\mathbf{q},0)\Upsilon
\bigg\rbrace
\label{funzione di partizione gaussiana temperatura nulla}
\end{equation}
with
\begin{equation}
\Omega_g^{(0)}= \frac{1}{2}\sum_{\mathbf{q}}E_q(\mu,\psi_0)
\label{granpotential gaussiano temperatura nulla}
\end{equation}
the zero-temperature Gaussian grand potential. At this point, by
recalling the relation $\Omega = \beta^{-1}\log\mathcal{Z}$, we are ready 
compute the quenched disorder average average  of the grand potential. It is immediate
to realize the $\Omega_{\text{mf}}$ and $\Omega_g^{(0)}$ are transparent to this
operation. On the contrary, from Eq. \eqref{funzione di partizione gaussiana temperatura nulla}
we easily derive that
\begin{equation}
\Omega_{\text{dis}}^{(0)}(U_{\text{D}}) = -\frac{1}{2}
\sum_{\mathbf{q}}\big|U_{\text{D}}(\mathbf{q})\big|^2
\Upsilon^T \mathcal{G}(\mathbf{q},0)\Upsilon
\label{granpotenziale di disordine temperatura nulla}
\end{equation}
whose quenched average reads
\begin{equation}
\braket{\Omega_{\text{dis}}}_{\text{th+dis}} =
-\frac{1}{2}\sum_{\mathbf{q}}\Upsilon^T\mathcal{G}(\mathbf{q},0)\Upsilon
\underbrace{\int \mathcal{D}[U_{\text{D}}] \,P[U_{\text{D}}] \;\big| 
	U_{\text{D}}(\mathbf{q})\big|^2}_{{}
	=\braket{|U_{\text{D}}(\mathbf{q})|^2}_{\text{dis}} = \gamma^2 }
\label{granpotenziale di disordine media quenched}
\end{equation}
thanks to Eq. \eqref{pdf per disordine delta correlato} (with only the spatial integration).
Thus, we have now isolated the disorder contribution to the zero-temperature
grand potential, namely
\begin{equation}
\begin{aligned}
\braket{\Omega_{\text{dis}}}_{\text{th+dis}} & =
- \frac{\gamma^2}{2}\sum_{\mathbf{q}} \Upsilon^T \mathcal{G}(\mathbf{q},0)\Upsilon \\
& = -\gamma^2 \psi_0^2 \sum_{\mathbf{q}} \frac{1}{\epsilon_q -\mu + 3g\psi_0^2}\;.
\end{aligned}
\label{granpotenziale di disordine solo somma momenti}
\end{equation}
Moving to the continuum limit, i.e. $\sum_{\mathbf{q}}\rightarrow \frac{L^d}{(2\pi)^d} 
\int d^d\mathbf{q}$, and performing the resulting integral with the aid of 
dimensional regularization, we finally obtain, in agreement with \cite{schakel-1997,schakel-book},
that
\begin{equation}
\frac{\braket{\Omega_{\text{dis}}^{(0)}}_{\text{th+dis}}}{L^d} =
- \frac{\Gamma(1-d/2)}{(2\pi)^{d/2}}\bigg(\frac{m}{\hbar^2}\bigg)^{d/2}
\gamma^2\psi_0^2\big(3g\psi_0^2 - \mu\big)^{(d-2)/2}\;
\label{granpotenziale di disordine finale}
\end{equation}
with $\Gamma(x)$ being the Euler Gamma function. From the equation above we can
compute the corresponding \textit{disorder} contribution to the total number density $n$
as a function of the condensate one $n_0$. Indeed, from 
$\braket{\Omega_{\text{dis}}}_{\text{dis}}$ in Eq. \eqref{granpotenziale di disordine finale}
we have
\begin{equation}
n_{\gamma}(n_0) = - \frac{1}{L^d}\frac{\partial \braket{\Omega_{\text{dis}}}_{\text{th+dis}}
(\mu,\psi_0)}
{\partial \mu}\bigg|_{\mu= g\psi_0^2 = g n_0} 
\label{densita di numero definizione}
\end{equation}
resulting in
\begin{equation}
n_{\gamma}(n_0) =  \frac{\Gamma(2-d/2)}{4\pi^{d/2}}\bigg(\frac{m}{\hbar^2}\bigg)^{d/2}
\;g^{d/2-2}\,n_0^{d/2-1} \gamma^2\;.
\label{contributo di disordine alla depletion}
\end{equation}
Concerning physical values of $d$, for $d=3$ we have $g=4\pi\hbar^2a_s/m$
with $a_s$ the s-wave scattering length. Consequently, one can verify that
\begin{equation}
\text{$d=3$}\quad \Longrightarrow \quad n_{\gamma}(n_0) = \frac{\gamma^2}{8\pi^{3/2}}
\bigg(\frac{m}{\hbar^2}\bigg)^2\,\sqrt{\frac{n_0}{a_s}}
\label{contributo di disordine alla depletion 3d}
\end{equation}
The two-dimensional case is much more complicated, since the
zero-range coupling constant displays a peculiar logarithmic dependence on the
number density \cite{boronat-2009,salasnich-2017,tononi-2018,tononi-2019}. For 
two-spatial dimensions a separate investigation is then needed, in order to investigate
the eventual divergences in the free energy and the relation between 
the disorder potential and the prediction of the Mermin-Wagner-Hohenberg theorem
\cite{lewenstein-2006,boudjema-2013,boudjema-2015}.
At $T=0$ the total number density is then given by
\begin{equation}
n = n_0 + n_g^{(0)}(n_0) + n_{\gamma}(n_0)\;,
\label{densita numero totale}
\end{equation}
with $n_g^{(0)}$ being the pure Gaussian contribution, derived from
Eq. \eqref{granpotential gaussiano temperatura nulla} as stated by 
Eq. \eqref{densita numero totale}. According to \cite{tononi-2018}
we have
\begin{equation}
n_g^{(0)}(n_0) = \frac{1}{4\pi^{(d+1)/2}} 
\frac{\Gamma\big(2-\frac{d}{2}\big)\Gamma\big(\frac{d-1}{2}\big)}
{\Gamma\big(\frac{d}{2}+1\big)}
\bigg(\frac{m}{\hbar^2}\bigg)^{d/2}
\big(g\,n_0\big)^{d/2}\;.  
\label{contributo gaussiano puro depletion temperatura nulla}
\end{equation}
It is crucial to remark that Eqs. \eqref{contributo di disordine alla depletion} and
\eqref{densita numero totale}
imply that, for a given number of total particles $n$, 
a smaller fraction of them
resides in the condensate. 
Thus, we can conclude that a disorder potential will cause an additional
depletion of the condensate. In order to extract the condensate fraction
$n_0/n$, Eq. \eqref{densita numero totale} has to be numerically solved.
Thus, for $d = 3$, by replacing Eq. \eqref{contributo gaussiano puro depletion temperatura nulla} 
and Eq. \eqref{contributo di disordine alla depletion 3d} in
Eq. \eqref{densita numero totale} we obtain
\begin{equation}
n = n_0 + \frac{8}{3\sqrt{\pi}} (n_0\, a_s)^{3/2}
+ \frac{\gamma^2}{8\pi^{3/2}}\bigg(\frac{m}{\hbar^2}\bigg)^2\, \sqrt{\frac{n_0}{a_s}}
\label{equazione di stato 3d}
\end{equation}
The equation above represents an implicit expression for the condensate fraction. At 
a given number density $n$, we can numerically solve it in $n_0$ to compute $n_0/n$.
Concerning Eq. \eqref{equazione di stato 3d} we have to underline another important detail.
It seems that the disorder contribution to $n$ 
(cfr. Eq. \eqref{contributo di disordine alla depletion 3d}) diverges in the very weakly
interacting limit $a_s\rightarrow 0$. However, the presence of a random external potential 
implies the presence of another characteristic length scale. Following \cite{falco-2007},
we define it in terms of the disorder correlator $\gamma$ given by 
Eq. \eqref{pdf per disordine delta correlato}, i.e. 
\begin{equation}
l_{\text{dis}} \equiv \frac{4\pi^2}{\gamma^2}\bigg(\frac{\hbar^2}{m}\bigg)^2 \;.
\label{lunghezza caratteristica del disordine}
\end{equation} 
Now, we recall that our perturbative
Gaussian scheme requires that $n_g^{(0)}$ 
in Eq. \eqref{contributo gaussiano puro depletion temperatura nulla} and
$n_{\gamma}$ from Eq. \eqref{contributo di disordine alla depletion} are both small
compared to $n_0\simeq n$. Thus, by imposing $n_g^{(0)}\ll n_0\simeq n$ we get the
condition $n^{1/3} a_s \ll 1$, the usual condition on the interaction strength.
At the same time, we have to require that $n_{\gamma} \ll n_0\simeq n$, leading us
$(n^{1/3}a_s)^{1/2} \, 2(n^{1/3}l_{\text{dis}})/\sqrt{\pi} \ll 1$. The latter 
highlights the fact that we have to require $n^{1/3}l_{\text{dis}} \gg 1$ in addition
to the usual condition on the gas parameter.
Provided that both the constrains hold, Eq. \eqref{equazione di stato 3d} can be 
perturbatively inverted, reading
\begin{equation}
\frac{n_0}{n} = 1- \frac{8}{3\sqrt{\pi}}\big(n\,a_s^3\big)^{1/2}
- \frac{\sqrt{\pi}}{2 (n^{1/3}l_{\text{dis}})\;\sqrt{n^{1/3}\,a_s}}\;.
\label{frazione condensata 3d}
\end{equation}
It is worth remembering at the end of this section that, in the case of quenched disorder,
the additional contribution to the grand potential is effectively a zero-temperature one
(or, in other words, it does not depend on $\beta$).
The reason is that, as we have stressed in the previous section, we are considering
an external random confinement whose time scale is much larger than the thermodynamic one.

\subsection{Superfluid response and quenched disorder}

Now, what can we say about the disorder influence on the superfluid motion
of the system?
Actually, within the Landau-Khalatnikov two-fluid formulation
\cite{landau-book,khalatnikov-book},
the calculation proceeds in a similar fashion, provided that the theory is modified
accordingly to the following points \cite{schakel-book,tononi-2018,tononi-2019}:
\begin{itemize}
	\item within the Landau two-fluid model, we assume that the normal part of the system
	is in motion with velocity $\mathbf{v}$, so $\partial_{\tau} \rightarrow \partial_{\tau}
	-i\mathbf{v}\cdot \nabla$;
	\item Superfluid and normal parts do not exchange momentum. This means that a superflow
	can be imposed by means of phase twist of the field $\psi(\mathbf{r},\tau)$, i.e.
	$\psi\rightarrow\exp\big( i m \mathbf{v}_s \cdot \mathbf{r}/\hbar \big)\psi $ with
	$\mathbf{v}_s$ the superfluid velocity \cite{fisher-1973,taylor-2006};
	\item  as a consequence of the previous points, we can redefine the chemical potential
	as 
	\begin{equation}
	\mu_{\text{eff}} = \mu - \frac{1}{2}m\mathbf{v}_s\cdot (\mathbf{v}_s - 2\mathbf{v})\;.
	\label{potenziale chimico efficace}
	\end{equation}
\end{itemize}
The resulting Gaussian partition function (or grand potential) 
is formally similar to the one in 
Eq. \eqref{funzione di partizione gaussiana fattorizzata fourier}, with the replacement
$\mu \rightarrow \mu_{\text{eff}}$ and the (inverse) Gaussian propagator given by
\begin{equation}
\mathcal{G}^{-1}(\mathbf{q},\omega)  = 
\begin{pmatrix}
-i\hbar\omega_n + \epsilon_q -\mu + 2g\psi_0^2 
+\hbar\mathbf{q}\cdot(\mathbf{v}-\mathbf{v}_s)& g\psi_0^2 \\
& \\
g\psi_0^2 & +i\hbar\omega_n + \epsilon_q -\mu + 2g\psi_0^2 
-\hbar\mathbf{q}\cdot(\mathbf{v}-\mathbf{v}_s)
\end{pmatrix}
\end{equation}
whose determinant is given by
\begin{equation}
\det\mathcal{G}^{-1}(\mathbf{q},\omega_n) = E_q^2 -
\big[i\hbar\omega_n-\hbar \cdot \mathbf{q}\cdot (\mathbf{v}-\mathbf{v}_s) \big]^2\;,
\label{determin}
\end{equation}
where $E_q$ as in Eq. \eqref{spettro delle fluttuazione sopra il punto sella}
with $\mu \rightarrow \mu_{\text{eff}}$.

 By following this prescription, the
\textit{phase-twisted} grand potential can be derived from 
Eq. \eqref{granpotenziale di disordine media quenched}, reading
\begin{equation}
\braket{\Omega_{\text{dis}}(\mu_{\text{eff}},\mathbf{v})}_{\text{th+dis}}=
- \gamma^2\psi_0^2 \sum_{\mathbf{q}}\frac{\epsilon_q-\mu_{\text{eff}}+
	g\psi_0^2}{E_q^2-\big[\hbar\mathbf{q}\cdot(\mathbf{v}-\mathbf{v}_s)\big]^2}\;.
\label{gran potenziale di disordine phase-twist}
\end{equation}
We are interested in the linear-response regime, where the density current
\begin{equation}
\mathbf{g} = - \frac{1}{L^d}\frac{\Omega(\mu_{\text{eff}}(\mathbf{v}),\mathbf{v})}
{\partial\mathbf{v}}
\label{corrente di densita}
\end{equation}
is taken up to the first order in $(\mathbf{v}-\mathbf{v}_s)$. 
Therefore, we begin with 
an expansion of Eq. \eqref{gran potenziale di disordine phase-twist} up to the
quadratic order in $\hbar\mathbf{q}\cdot (\mathbf{v}-\mathbf{v}_s)$. For symmetry reasons
the linear term of this expansion is identically zero, such that we
have to deal with the following equation, 
\begin{equation}
\frac{\braket{\Omega_{\text{dis}}(\mu_{\text{eff}},\mathbf{v})}_{\text{th+dis}}}{L^d}=
- \frac{\gamma^2\psi_0^2}{(2\pi)^d} \int d^d\mathbf{q}\bigg[\frac{\epsilon_q-\mu_{\text{eff}}+
	g\psi_0^2}{E_q^2}+\frac{\hbar^2q^2}{d} \bigg(\frac{\epsilon_q-\mu_{\text{eff}}+
	g\psi_0^2}{E_q^4}\bigg)(\mathbf{v}-\mathbf{v}_s)^2
\bigg]\;,
\label{granpotenziale phase twist 2}
\end{equation}
where we have made use of the identity 
\begin{equation}
\int d^d\mathbf{q}\, q_iq_j f(|\mathbf{q}|) = d^{-1}\delta_{ij}\int d^d\mathbf{q}
\,|\mathbf{q}|^2f(|\mathbf{q}|)\;.
\end{equation}
Since the first term in the equation above depends only on $\mu_{\text{eff}}$
and it holds $\frac{\partial \mu_{\text{eff}}}{\partial \mathbf{v}} = m\mathbf{v}_s$, 
it will be the disorder contribution to the $n\mathbf{v}_s$ term in $\mathbf{g}$. More
precisely, by considering also the thermal depletion due to the quasiparticle excitations,
Eq. \eqref{corrente di densita} reads
\begin{equation}
\mathbf{g} = n\mathbf{v}_s + \frac{1}{d}\int\frac{d^d\mathbf{q}}{(2\pi)^d}
\bigg[ 
\frac{\beta\hbar^2 q^2}{m} \frac{e^{\beta E_q}}{(e^{\beta E_q} - 1)^2}
+ 4\gamma^2\psi_0^2 \frac{\epsilon_q(\epsilon_q-\mu+g\psi_0^2)}{E_q^4}
\bigg](\mathbf{v}-\mathbf{v}_s)\;.
\label{corrente densita risultato finale}
\end{equation}
At this point, we finally replace Eq. \eqref{saddle point} for the 
broken-symmetry case in the equation above. A simple comparison between
the resulting equation and the the two-fluid assumption 
$\mathbf{g}= n_s\mathbf{v}_s + n_n\mathbf{v}$ leads us to
\begin{equation}
n_n(n_0,T) = \frac{\beta}{d}\int \frac{d^d\mathbf{q}}{(2\pi)^d}\frac{\hbar^2 q^2}{m}
\frac{e^{\beta E_q}}{(e^{\beta E_q} -1)^2} + \frac{4}{d}n_{\gamma}(n_0)
\label{superfluid depletion}
\end{equation}
with $n_{\gamma}(n_0)$ as in Eq. \eqref{contributo di disordine alla depletion}.
\begin{figure}[ht!]
	\includegraphics[width=0.85\columnwidth,clip=]{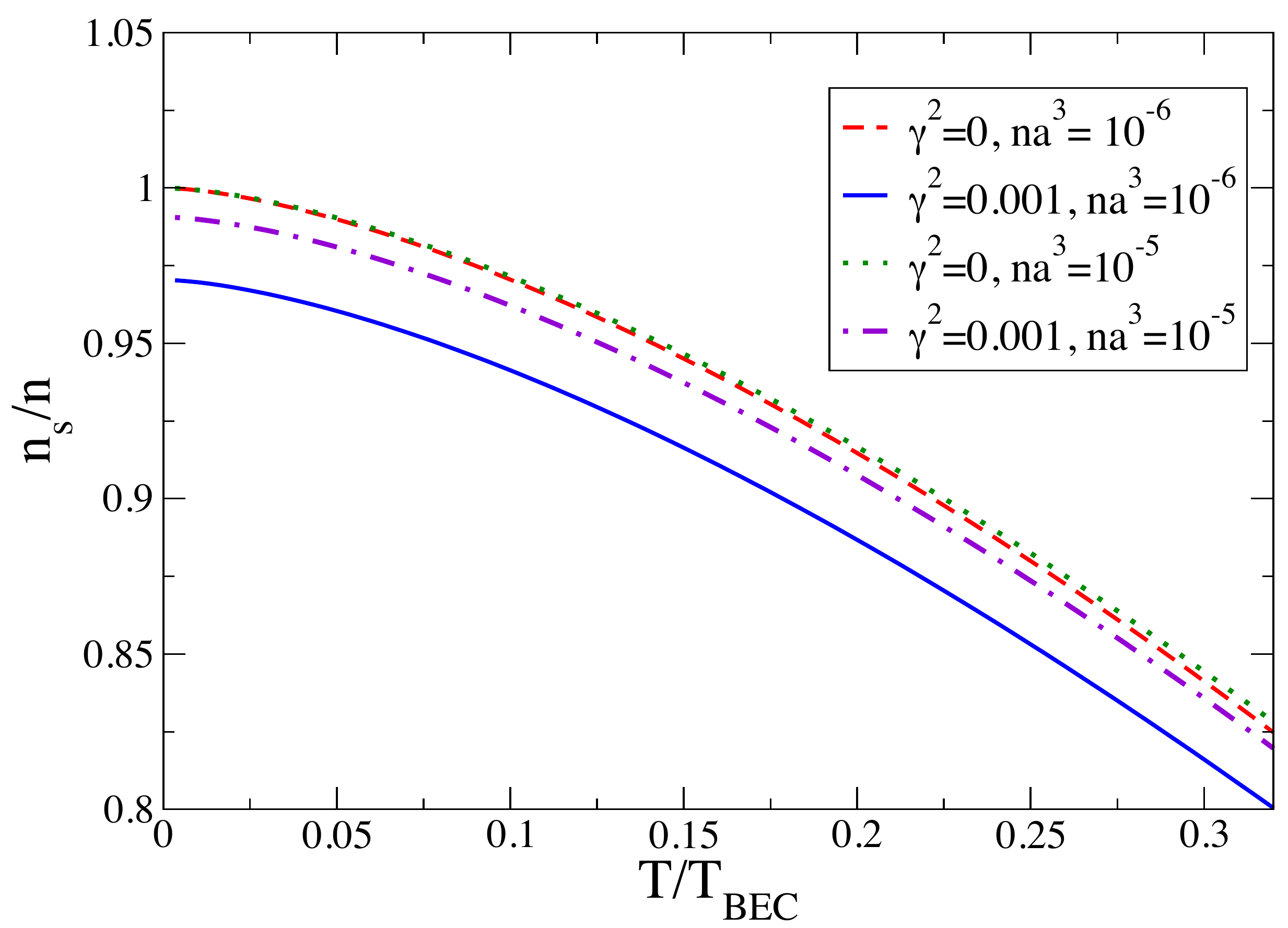}
	\caption{Plot of superfluid fraction in presence and absence of an uncorrelated
		disordered potential. The curve are derived from the two-fluid assumption
		$n=n_s + n_n$, with the normal contribution $n_n$ given by Eq. \eqref{superfluid depletion}.
		The temperatures are in units of $T_{\text{BEC}} = 2\pi \hbar^2 n^{2/3}
		/(mk_B\zeta(3/2)^{2/3})$, i.e. the condensation temperature of an ideal Bose gas.
		The solid blue line and the dot-dashed violet one represent the behaviour of $n_s/n$ 
		in presence of a disordered potential with $\gamma^2 = 0.001$ for two different
		values of the gas parameter $na_s^3 = 10^{-6}$ and $10^{-5}$ respectively. 
		For the same values of $na_s^3$, the dashed red line and the dotted green 
		correspond the case where there is no disorder contribution ($\gamma^2 = 0$).}
	\label{fig:1}
\end{figure}
In order to understand the relevance of Eq. \eqref{superfluid depletion}, one has to remember
that, in the pure system (i.e. $\gamma = 0$), all the particles belong to the superflow.
On the contrary, a disorder potential induce a superfluid depletion also at $T = 0$ and,
in addition, the parameter $\gamma$ may act as knob to destroy the superflow since
\begin{equation}
n_s^{(0)}(n_0) = n - \frac{4}{d}n_{\gamma}(n_0)\;.
\label{frazione superfluida svuotata dal disordine}
\end{equation} 
In a similar way to the condensate fraction, we are not allowed to expand
$n_s/n$ in the limit $g\rightarrow 0$ because
Eq. \eqref{contributo di disordine alla depletion} is divergent in that regime.
Obviously, the superfluid density can be computed from the two-fluid assumption
$n=n_s+n_n$ where $n_n$ is given by Eq. \eqref{superfluid depletion}.

In fig. \ref{fig:1} we report the behaviour of the superfluid fraction in presence of a
disordered but uncorrelated external potential with $\gamma^2 = 0.001$ for  
two different values of the gas parameter $na_s^3$. In order to highlight the
crucial contribution due to disorder, we also consider the case where $\gamma^2=0$. 
It is immediate to realize that in pure system 
($\gamma^2=0$) the whole system is superfluid at $T=0$, while the presence of a disorder
potential lowers the curves, since there is an additional depletion at $T = 0$.
It is also interesting to underline that the more dilute is the system, the 
more evident is the detachment from the pure system.

\section{The replicated action for superfluid bosons}
\label{sec:4}

The previous section has been focused on the extension of usual field-theory
techniques superfluid bosons subject to a random external confinement. Moreover,
the whole calculation has been carried out in the quenched regime, where we effectively average
over different microscopic realization of the system with a given disorder
configuration. 

We have already made clear that this is not the only viable strategy and that it does not
handle physical realizations where, for instance, the disorder potential also has a thermal 
component. This means that, looking at Eq. \eqref{media con il metodo delle sorgenti}, it is not 
clear in which order integration and differentiation have to be performed. In order to overcome
this ambiguity, the replica trick represents one of the most ingenious strategy. It was
first developed in the context of spin glasses 
\cite{anderson-1975,parisi-book,parisi-review} but, in the following, we are going to detail
how it can be applied to describe the role of disorder in ultracold bosonic gases.

Let us begin by considering again Eq. \eqref{media con il metodo delle sorgenti}. 
The following three hypothesis represent the starting of the replica formalism. Thus,
\begin{itemize}
	\item We assume that $\mathcal{Z}[J=0] = 1$, establishing a proper normalization and implying
	that Eq. \eqref{media con il metodo delle sorgenti} has no dependence on $V_{\text{D}}$ in the
	denominator;
	\item  it holds $\braket{\mathcal{O}}_{\text{th+dis}} = - 
	\delta\braket{\log\mathcal{Z}}_{\text{dis}}/\delta_J|_{J=0}$;
	\item for $J\neq 0$ the partition function retains its algebraic properties.
\end{itemize}
Actually, the last point is a somewhat obscure way to state that the
average procedure involves functionals linear in the disorder potential, since
$V_{\text{D}}$ has been linearly added to the action. 
The great advantage provided by the replica trick can be understood from the
formal relations reported below,
\begin{equation}
\log\mathcal{Z} [J] = \lim_{R\rightarrow 0} \frac{e^{R\log\mathcal{Z}}-1}{R}
= \lim_{R\rightarrow 0}\frac{\mathcal{Z}^R - 1}{R}
\label{vantaggio del replica trick}
\end{equation}
with $R \in\mathbb{N}$. The replacement of $\log\mathcal{Z}$ with an 
algebraic function of $\mathcal{Z}$ strongly reduces the complexity. Indeed,
concerning the calculation of the expectation values, we have
\begin{equation}
\braket{\mathcal{O}}_{\text{th+dis}} = -\lim_{R\rightarrow 0} \frac{\delta}{\delta J}
\frac{\mathcal{Z}^R}{R}\;.
\label{media con il replica trick}
\end{equation}
Actually, within the replicated formalism, expectation values can be obtained by simply
considering $R$ identical copies of the system under consideration. Concerning the
functional integration, it does not alter the linear dependence of the action on the
disorder field, because $[\exp(\int dV_{\text{D}})]^R = \exp(\sum^{R}\int dV_{\text{D}})$.
Unfortunately all this comes with a price. Looking at Eq. \eqref{media con il replica trick}
we realize that, at the end, we should perform an analytical continuation $R\rightarrow 0$.
The point is that there is no guarantee at all that the function 
$\delta_J\braket{\mathcal{Z}^R}_{\text{dis}}$ is analytic all the way down to $R = 0$
\cite{altland-book}. This means that, from a mathematical standpoint, the replica trick 
is not a well-founded technique. Surprisingly, it is also true that it seems to rarely fail, 
at least when the disorder can be treated in a perturbative way 
\cite{parisi-book,parisi-review}.
\\

Let us consider in the following the generalization of 
Eq. \eqref{pdf per disordine delta correlato}, namely 
given by
\begin{equation}
\tilde{\mathsf{P}}[U_{\text{D}}] = \exp\bigg\lbrace
- \frac{1}{2}\int_0^{\beta\hbar}d\tau \int d^d\mathbf{r}
\;\gamma^{-2}(\mathbf{r},\tau) U_{\text{D}}^2(\mathbf{r},\tau)
\bigg\rbrace
\label{pdf per disordine generalizzata}
\end{equation}
where, actually, the disorder is assumed translationally invariant in time and space.
In order to implement the replica trick within the context of superfluid bosonic
systems, as an alternative to Eq. \eqref{splitting del campo 1} we move to the phase-density
representation, where
\begin{equation}
\psi(\mathbf{r},\tau) = \sqrt{n_0+\pi(\mathbf{r},\tau)}\;e^{i\varphi(\mathbf{r},\tau)} \;.
\label{splitting del campo densita-fase}
\end{equation}
with $n_0 = \psi_0^2$, $\psi_0^2$ being given by Eq. \eqref{saddle point}.
From the equation above we deduce that 
$n(\mathbf{r},\tau) = n_0+\pi(\mathbf{r},\tau)$ and a perturbative strategy is easily 
implemented by assuming the smallness of density fluctuation. 
Thus, Eq. \eqref{splitting del campo densita-fase} becomes
\begin{equation}
\psi(\mathbf{r},\tau) \simeq \sqrt{n_0}\bigg[1+\frac{\pi(\mathbf{r},\tau)}{2n_0} \bigg]
e^{i\varphi(\mathbf{r},\tau)}\;.
\label{density-phase piccole fluttuazioni}
\end{equation}
In order to understand the limitations of our approach, it is crucial to recall that
$\braket{n(\mathbf{r},\tau)}_{\text{th}} = \braket{|\psi(\mathbf{r},\tau)|^2}_{\text{th}}$,
while we often set $n \simeq n_0$ with 
$n_0 = \big|\braket{\psi(\mathbf{r},\tau)}_{\text{th}}\big|^2$. Up to the
Gaussian level this is an acceptable approximation since 
$\braket{\pi(\mathbf{r},\tau)}_{\text{th}} \simeq 0$.
In the regime of small fluctuations, we can expand the mean-field analysis by retaining 
only the quadratic terms in $\pi(\mathbf{r},\tau)$ and $\varphi(\mathbf{r},\tau)$.
The change of variables $\lbrace\psi,\psi^* \rbrace\rightarrow \lbrace\pi,\varphi\rbrace$ 
has a constant Jacobian determinant, therefore Eq. \eqref{funzione di partizione grancanonica}
reads
\begin{equation}
\mathcal{Z}\simeq \int \mathcal{D}[\pi,\varphi]
\exp\bigg\lbrace-\frac{1}{\hbar} S_g[\pi,\varphi]\bigg\rbrace
\label{funzione di partizione grancanonica gaussiana}
\end{equation}
with the periodic boundary conditions 
$\pi(\mathbf{r},\beta\hbar) = \pi(\mathbf{r},0)$ and
$\varphi(\mathbf{r},\beta\hbar) = \varphi(\mathbf{r},0)$. The smallness
of fluctuations allows to extend the integration range of the phase variable
from $[-\pi,+\pi]$ to $[-\infty,+\infty]$. This decompactification has obviously to be
addressed with particular attention when dealing with $d\leq 2$, where the enhancement of
fluctuations prevents the occurring of a \textit{true} long-range order and a more
refined analysis is required 
\cite{altland-book,mermin-1966,hohenberg-1967,villain-1975,kadanoff-1977}.
The Gaussian action in  
Eq. \eqref{funzione di partizione grancanonica gaussiana} is given by
\begin{equation}
\begin{aligned}
S_g[\pi,\varphi] &= \int_{0}^{\beta\hbar} d\tau\int d^d\mathbf{r}
\bigg[i\hbar\pi\frac{\partial \varphi}{\partial \tau}
+\frac{n_0\hbar^2 (\nabla \varphi)^2}{2m}
+ \frac{\hbar^2(\nabla\pi)^2}{8mn_0} -U_{\text{D}}(\mathbf{r},\tau)
(n_0+ \pi) \\
& \qquad \qquad + \frac{1}{2}\int d^d\mathbf{r}'\pi(\mathbf{r}',\tau)
V(|\mathbf{r}-\mathbf{r}'|)\pi(\mathbf{r},\tau)
\bigg] 
- \frac{1}{2}\beta\hbar L^d \tilde{V}(0)n_0^2
\end{aligned}
\label{azione campo medio con gaussiana}
\end{equation}
where we have imposed Eq. \eqref{saddle point} to eliminate the dependence on the
chemical potential and $U_{\text{D}}$ is distributed according to 
Eq. \eqref{pdf per disordine generalizzata}. 
The pertubative expansion up to the terms contained
in Eq. \eqref{azione campo medio con gaussiana} is equivalent to a one-loop expansion
within a diagrammatic representation. Equivalently, in the second-quantization
formalism this corresponds to the Bogoliubov approximation \cite{andersen-2004}. 

The implementation of the replica trick requires the calculation
of $\braket{\mathcal{Z}^R}_{\text{dis}}$. The (not-averaged)  partition
function is simply
\begin{equation}
\mathcal{Z}^R = \int \mathcal{D}[\pi,\varphi] \exp\bigg\lbrace
- \frac{1}{\hbar}\sum_{\alpha=1}^R S_g[\pi_{\alpha},\varphi_{\alpha}]
\bigg\rbrace\;,
\label{funzione di partizione con le repliche non mediata}
\end{equation}
where $S_g[\pi_{\alpha},\varphi_{\alpha}]$ is provided by
Eq. \eqref{azione campo medio con gaussiana}
and $\mathcal{D}[\pi,\varphi] = \prod_{\alpha=1}^R \mathcal{D}[\pi_{\alpha},\varphi_{\alpha}]$. 
In case of
a Gaussian disorder obeying to Eq. \eqref{pdf per disordine generalizzata}, 
Eq. \eqref{disorder average con pdf} reads
\begin{equation}
\begin{aligned}
\braket{\mathcal{Z}^R}_{\text{dis}} & = \int \mathcal{D}[\pi,\varphi] \exp\bigg\lbrace
- \frac{1}{\hbar}\sum_{\alpha=1}^R S^{\text{(pure)}}_g[\pi_{\alpha},\varphi_{\alpha}]
\bigg\rbrace \times \\
& 
\times \underbrace{\int \mathcal{D}[U_{\text{D}}]\exp\bigg\lbrace 
	- \frac{1}{2}\int_0^{\beta\hbar}d\tau\int d^d\mathbf{r}
	 \;\gamma^{-2}(\mathbf{r},\tau)\,U_{\text{D}}^2(\mathbf{r},\tau)
	\bigg\rbrace \exp\bigg\lbrace 
	- \frac{1}{\hbar} \sum_{\alpha=1}^R S^{\text{(coup)}}_g[U_{\text{D}},\pi_{\alpha}]
	\bigg\rbrace}_{{}=(\star)}
\end{aligned}
\label{funzione di partizione replicata e mediata 1}
\end{equation}
where $S^{\text{(pure)}}_g[\pi_{\alpha},\varphi_{\alpha}]$ 
is Eq. \eqref{azione campo medio con gaussiana} with 
$U_{\text{D}} = 0$ and $S^{\text{(coup)}}_g[U_{\text{D}},\pi_{\alpha}]$
is given by
\begin{equation}
S^{\text{(coup)}}_g[U_{\text{D}},\pi_{\alpha}] =
\int_0^{\beta\hbar}d\tau\int d^d\mathbf{r}\; U_{\text{D}}(\mathbf{r},\tau)
\sum_{\alpha=1}^R \pi_{\alpha}(\mathbf{r},\tau)\;.
\label{coupling disordine fluttuazione densita}
\end{equation}
The second line of Eq. \eqref{funzione di partizione replicata e mediata 1}
is a Gaussian integral in $U_{\text{D}}$ such that
\begin{equation}
(\star) = \exp \bigg\lbrace 
\frac{1}{2\hbar} \int_0^{\beta\hbar} d\tau \int d^d\mathbf{r}\; 
\gamma^2(\mathbf{r},\tau)
\sum_{\alpha,\beta}
\pi_{\alpha}(\mathbf{r},\tau)\pi_{\beta}(\mathbf{r},\tau)
\bigg\rbrace\;.
\label{integrale gaussiano per partizione replicata}
\end{equation} 
In this way, the replicated partition in 
Eq. \eqref{funzione di partizione replicata e mediata 1}
results in 
\begin{equation}
\braket{\mathcal{Z}^R}_{\text{dis}} = \int \mathcal{D}[\pi,\varphi] \exp\bigg\lbrace
-\frac{1}{\hbar}\sum_{\alpha=1}^R S_g^{\text{(pure)}} [\pi_{\alpha},\varphi_{\alpha}]
-\frac{1}{\hbar}\sum_{\alpha,\beta = 1}^RS_g^{\text{(dis)}} 
[\pi_{\alpha},\pi_{\beta},\varphi_{\alpha},\varphi_{\beta}]
\bigg\rbrace\;.
\label{funzione di partizione replicata e mediata fin}
\end{equation}
For the sake of clarity we report below $S_g^{\text{(pure)}}$
and $S_g^{\text{(dis)}} $, namely
\begin{align}
& S_g^{\text{(pure)}} [\pi_{\alpha},\varphi_{\alpha}] =
\int d^{d+1}\mathbf{s} \bigg[ 
i\hbar\pi_{\alpha}\partial_{\tau}\varphi_{\alpha} + 
\frac{n_0\,\hbar^2(\nabla\varphi_{\alpha})^2}{2m}
+\frac{\hbar^2(\nabla\pi_{\alpha})^2}{8m n_0} + \frac{1}{2}
\int d^d\mathbf{r}'\pi(\mathbf{r}',\tau)
V(|\mathbf{r}-\mathbf{r}'|)\pi(\mathbf{r},\tau)
\bigg] 
\label{azione del sistema puro}\\
& S_g^{\text{(dis)}}
[\pi_{\alpha},\pi_{\beta},\varphi_{\alpha},\varphi_{\beta}] =
- \frac{1}{2} \int d^{d+1}\mathbf{s}\, \gamma^2(\mathbf{r},\tau)\pi_{\alpha}(\mathbf{s})
\pi_{\beta}(\mathbf{s})
\label{contributo di disordine alla azione replicata}
\end{align}
with $\mathbf{s}= (\mathbf{r},\tau)$. It is crucial to underline the relevance
of Eq. \eqref{funzione di partizione replicata e mediata fin}. Indeed, it shows that
also in presence of the simplest disorder potential ($\delta$-correlated both in time and
space), the outlined formalism generates an effective quartic interaction between different
replica indices. This fact can be understood by recalling the physical meaning of Feynman's path.
Despite its complexity, and indifferent to different replicas, the path integrals prefer
to evolve towards the lowest possible potential. In other words they have a tendency
to populate the same regions of the energy landscape.

Moving to the calculation of expectation values, the replicated formalism 
replaces the logarithm of the partition function with, basically, 
Eq. \eqref{funzione di partizione replicata e mediata fin}, where the Euclidean action 
of Eq. \eqref{azione campo medio con gaussiana} is splitted into 
Eq. \eqref{azione del sistema puro} and Eq. \eqref{contributo di disordine alla azione replicata}.
According to Eq. \eqref{media con il replica trick}, the joint thermal and
disorder average is obtained through
\begin{equation}
\braket{{\,\mathcal{O}\,}}_{\text{th+dis}} = -\lim_{R\rightarrow 0} 
\frac{\delta}{\delta J} \frac{1}{R}\mathcal{Z}^R[J] = \lim_{R\rightarrow 0}
\frac{1}{R}
\braket{\,\mathcal{O}[\psi_{\alpha},\psi^*_{\alpha}]\,}_{\text{rep}}
\label{media termica e disordine con le repliche}
\end{equation}
where $\braket{\bullet}_{\text{rep}}$ has to be intended as a thermal average over the
replicated partition function in Eq. \eqref{funzione di partizione replicata e mediata fin} or,
equivalently, over the action given by the sum of Eq. \eqref{azione del sistema puro} and
Eq. \eqref{contributo di disordine alla azione replicata}, i.e.
\begin{equation}
\braket{\,\mathcal{O}\,}_{\text{rep}} =
\dfrac{\int\mathcal{D}[\pi,\varphi] \mathcal{O}
	\exp\lbrace\hbar^{-1}( S_g^{\text{(pure)}}[\pi,\varphi]+S_g^{\text{(dis)}}[\pi,\varphi]) 
	\rbrace}
{\int\mathcal{D}[\pi,\varphi]\exp\lbrace\hbar^{-1}( S_g^{\text{(pure)}}[\pi,\varphi]+S_g^{\text{(dis)}}[\pi,\varphi])\rbrace }\;.
\label{media termica replicata}
\end{equation}

It is useful to expande the fields in Fourier series
such that
Eq. \eqref{azione del sistema puro} reads
\begin{equation}
S_g^{\text{(pure)}}[\pi_{\alpha},\varphi_{\alpha}]=
\frac{1}{2\beta\hbar L^d}\sum_{\mathbf{q}}\sum_n \chi_{\alpha}^T(-\mathbf{q},-\omega_n)
\mathcal{M}^{-1}(\mathbf{q},\omega_n)\chi_{\alpha}(\mathbf{q},\omega_n)
\label{azione del sistema puro fourier}
\end{equation}
where we have defined the vector $\chi_{\alpha}$ as
$
\chi_{\alpha}(\mathbf{q},\omega_n)^T \equiv
\big(
\varphi_{\alpha}(\mathbf{q},\omega_n);
\pi_{\alpha}(\mathbf{q},\omega_n)
\big)
$
The pure inverse propagator in Eq. \eqref{azione del sistema puro fourier}
is given by
\begin{equation}
\mathcal{M}^{-1}(\mathbf{q},\omega_n) = 
\begin{pmatrix}
n_0\dfrac{\hbar^2 q^2}{m} & & -\hbar\omega_n \\
+\hbar\omega_n & & \dfrac{\hbar^2 q^2}{4mn_0} + \tilde{V}(\mathbf{q}) 
\end{pmatrix}\;.
\label{propagatore gaussiano del sistema pure}
\end{equation}
On the other hand, for Eq. \eqref{coupling disordine fluttuazione densita} we have now
\begin{equation}
S_g^{\text{(dis)}}[\pi_{\alpha},\pi_{\beta},\varphi_{\alpha},\varphi_{\beta}]=
-\frac{1}{2\beta\hbar L^d} \sum_{\mathbf{q}}\sum_n \gamma^2(\mathbf{q},\omega_n)
\pi_{\alpha}^*(\mathbf{q},\omega_n)\pi_{\beta}(\mathbf{q},\omega_n)\;.
\label{contributo di disordine fourier}
\end{equation}
The equation above can be decomposed in its diagonal and off-diagonal contributions, leading 
us to the final equation for the Euclidean action of Gaussian fluctuations. By renaming
$Q = (\mathbf{q},\omega_n)$ and using the fact that $x(-Q) = x^*(Q)$, we have
\begin{equation}
S_g^{\text{(rep)}}[\pi,\varphi] = \frac{1}{2\beta\hbar L^d} \sum_{Q} \mathsf{X}^{\dagger}(Q) \mathbb{M}^{-1}(Q) 
\mathsf{X}(Q)\;.
\label{azione gaussiana replicata e compatta}
\end{equation}
In the equation above, the sum over the replicas is encoded in the
$(R\times 2)$-dimensional column vector 
\begin{equation}
\mathsf{X}(Q) = 
\begin{pmatrix}
\chi_{1}(Q) \\
\chi_2(Q) \\
\vdots \\
\chi_R(Q)
\end{pmatrix}\;.
\label{vettore colonna esteso alle repliche}
\end{equation}
The 
$R\times 2$ matrix $\mathbb{M}^{-1}(Q)$ in Eq. \eqref{azione gaussiana replicata e compatta}
is built as
\begin{equation}
\mathbb{M}^{-1}(Q) = 
\begin{pmatrix}
\tilde{\mathcal{M}}^{-1}(Q) & \mathbb{B}(\gamma) & \dots & \dots & \mathbb{B}(\gamma) \\
\mathbb{B}(\gamma) & \tilde{\mathcal{M}}^{-1}(Q) & \dots & \dots & \mathbb{B}(\gamma) \\
\vdots & \vdots & \ddots &  & \vdots \\
\vdots & \vdots & &\ddots  & \mathbb{B}(\gamma) \\
\mathbb{B}(\gamma) & \mathbb{B}(\gamma) & \dots & \mathbb{B}(\gamma) & \tilde{\mathcal{M}}^{-1}(Q)
\end{pmatrix}
\label{propagatore gaussian circolante fourier}
\end{equation}
with the generating blocks
\begin{align}
\tilde{\mathcal{M}}^{-1}(Q) &= 
\begin{pmatrix}
n_0\frac{\hbar^2 q^2}{m} & & -\hbar\omega_n \\
+\hbar\omega_n & & \frac{\hbar^2 q^2}{4mn_0} + \tilde{V}(\mathbf{q})-
\gamma^2 (\mathbf{q},\omega_n)
\end{pmatrix}
\label{blocco generatore diagonale del propagatore}\\
& \notag\\
\mathbb{B}(\gamma) &= 
\begin{pmatrix}
0 & 0 \\
0 & -\gamma^2(\mathbf{q},\omega_n) \\
\end{pmatrix}\;.
\label{blocco generatore fuori diagonale del propagatore}
\end{align}

\section{Correlations functions in the replicated formalism}
\label{sec:5}

As detailed in App. \ref{app:A1}, the matrix $\mathbb{M}$ 
in Eq. \eqref{propagatore gaussian circolante fourier} 
is diagonalized by a well-known unitary transformation, reading
\begin{equation}
\tilde{\mathbb{M}}^{-1}(Q) = \text{diag}\big\lbrace
\tilde{\mathcal{M}}_R^{-1}(Q),\underbrace{\tilde{\mathcal{M}}^{-1}(Q), \ldots,
\tilde{\mathcal{M}}^{-1}(Q)}_{R-1 \quad \text{times}}
\big\rbrace
\label{propagatore replicato diagonalizzato}
\end{equation}
with
\begin{equation}
\mathbb{M}^{-1}_R(Q) = 
\begin{pmatrix}
n_0 \frac{\hbar^2 q^2}{m} & & -\hbar\omega_n \\
+\hbar\omega_n & &\frac{\hbar^2 q^2}{4mn_0} +\tilde{V}(\mathbf{q})-R\gamma^2(\mathbf{q},\omega_n)
\end{pmatrix}\;.
\end{equation}
Up to this level of approximation,
correlation functions are specified by the entries of the fluctuation propagator. 
For the pure system, by assuming translational invariance both in time and space,
we have 
\begin{equation}
\braket{\chi_j(\mathbf{q},\omega_n)\chi_k(\mathbf{q}',\omega_{n'})}_{\text{th}} = 
\mathcal{M}_{jk}
(\mathbf{q},\omega_{n})\delta(\mathbf{q}+\mathbf{q}')\beta\hbar\delta_{n,-n'}\;.
\label{valori di aspettazione per il sistema pure}
\end{equation}
On the other hand, according to Eq. \eqref{media termica e disordine con le repliche},
the replicated formalism provides both the disorder and the thermal average at the same
time. For instance, concerning the density-density correlation function, we have
\begin{equation}
\begin{aligned}
\braket{\pi^*(\mathbf{q},\omega_{n})\pi(\mathbf{q}',\omega_{n'})}_{\text{dis+th}} & = 
\lim_{R\rightarrow 0} \frac{1}{R}\sum_{\alpha=1}^R
\braket{\pi_{\alpha}(\mathbf{q},\omega_n) \pi_{\alpha}(\mathbf{q}',
	\omega_{n'})}_{S_g^{\text{(rep)}}}\\
& = \lim_{R\rightarrow 0}\frac{1}{R}\bigg[
(R-1)\frac{n_0 \hbar^3q^2/m}{\hbar^2\omega_{n}^2 + E_q^2} + 
\frac{n_0 \hbar^2q^2/m}{\hbar^2\omega_{n}^2 + E_q^2 -R\gamma^2(\mathbf{q},\omega_n)
n_0 \dfrac{\hbar^2 q^2}{m}}
\bigg]\delta(\mathbf{q}+\mathbf{q}')\beta\hbar\delta_{n,-n'} \\
& = \bigg[
\frac{n_0\hbar^3q^2/m}{\hbar^2\omega_{n}^2+E_q^2} + 
\gamma^2(\mathbf{q},\omega_{n})
\frac{n_0\hbar^2q^2q'^2/m}{\big(\hbar^2\omega_{n}^2 + E_q^2\big)
\big(\hbar^2\omega_{n'}^2 +E_{q'}\big)}
\bigg]\delta(\mathbf{q}+\mathbf{q}')\beta\hbar\delta_{n,-n'}
\end{aligned}
\label{correlazione gaussiana density-density}
\end{equation}
where the final result is obtained by expanding the second line up to the first
order in $R$. Moreover, we have
\begin{equation}
E_q = \sqrt{\epsilon_q\big[\epsilon_q + 2n_0\tilde{V}(\mathbf{q})\big]}\;,
\end{equation}
nothing else than Eq. \eqref{spettro delle fluttuazione sopra il punto sella}
with the saddle-point replacement $\mu = \tilde{V}(0)\,n_0$.
The remaining correlations are computed in the same way, leading to
\begin{equation}
\begin{aligned}
\braket{\varphi^*(\mathbf{q},\omega_{n})\varphi(\mathbf{q}',\omega_{n'})}_{\text{dis+th}} & =
\bigg[
\frac{mE_q^2/(n_0\hbar q^2)}{\hbar^2\omega_{n'}^2+E_q^2} 
 + \frac{\hbar^2\omega_{n}\omega_{n'}}
{\big(\hbar^2\omega_{n}^2 + E_q^2\big)
	\big(\hbar^2\omega_{n'}^2 +E_{q'}\big)}\gamma^2(\mathbf{q},\omega_{n})
\bigg]\delta(\mathbf{q}+\mathbf{q}')\beta\hbar\delta_{n,-n'} \\
\braket{\varphi^*(\mathbf{q},\omega_{n})\pi(\mathbf{q}',\omega_{n'})}_{\text{dis+th}} & =
\bigg[
\frac{\hbar^2\omega_{n}}{\hbar^2\omega_{n}^2 + E_q^2} +
\frac{ n_0\hbar^3 q'^2\omega_{n'}/m}
{\big(\hbar^2\omega_{n}^2 + E_q^2\big)
	\big(\hbar^2\omega_{n'}^2 +E_{q'}\big)}\gamma^2(\mathbf{q},\omega_{n})
\bigg]\delta(\mathbf{q}+\mathbf{q}')\beta\hbar\delta_{n,-n'}
\end{aligned}
\label{correlazioni gaussiane fase-fase e fase-densita}
\end{equation}
Besides the phase and density correlators, one of the most important
correlation functions is the current one, defined as
\begin{equation}
\mathcal{C}_{ij}(\mathbf{r},\tau;\mathbf{r}',\tau') = m^{-1}
\big[\braket{ g_i(\mathbf{r},\tau)g_j(\mathbf{r}',\tau')}
-\braket{g_i(\mathbf{r},\tau)}
\braket{g_j(\mathbf{r}',\tau')}
\big]\;,
\label{densita di corrente definizione}
\end{equation}
where the notation $\braket{\ldots}$ represents, for now, a general average procedure.
In the phase-density representation, from Eq. \eqref{splitting del campo densita-fase}
it is easy to verify that the current density $\mathbf{g} = \hbar (n_0+\pi)\nabla\varphi/m$.
Thus, by making use of the Wick theorem \cite{fetter-book} to disentagle the four-field
correlator in Eq. \eqref{densita di corrente definizione}, in the Fourier space one obtains
\cite{tauber-1997}
\begin{equation}
\begin{aligned}
\mathcal{C}_{ij}(\mathbf{q},\omega_n;\mathbf{q}',\omega_{n'}) & =
-\hbar^2 m n_0^2 q_i q_j\Braket{\varphi(\mathbf{q},\omega_n)
	\varphi(\mathbf{q}', \omega_{n'})} + \\
& - \frac{1}{(\beta L^d)^2}
\sum_{\mathbf{k},\omega_m}\sum_{\mathbf{k}',\omega_{m'}}k_i k'_j \bigg[
\Braket{\pi(\mathbf{q}-\mathbf{k}, \omega_{l-m}) 
	\pi(\mathbf{q}'-\mathbf{k}',\omega_{l'-m'})
}
\Braket{\varphi(\mathbf{k},\omega_m)\varphi(\mathbf{k}',\omega_{m'})} + \\
&\qquad \qquad  + \Braket{\pi(\mathbf{k}-\mathbf{q},\omega_{m-l}) \varphi(\mathbf{k}',\omega_{m'})}
\Braket{\pi(\mathbf{k}'-\mathbf{q}',\omega_{m'-l'})\varphi(\mathbf{k},\omega_m)}
\bigg]\;.
\end{aligned}
\label{correlator in fourier space after wick}
\end{equation}
Thanks to the replica trick, the disorder contribution is taken into account by
considering the correlators over the replicated Gaussian action, namely
Eqs. \eqref{correlazione gaussiana density-density} and 
\eqref{correlazioni gaussiane fase-fase e fase-densita}.

The current response function in Eq. \eqref{correlator in fourier space after wick} 
is crucial since it encodes the superfluid character of the system. 
This point is
clarified by noticing that $\mathcal{C}_{ij}$ is a 2-rank tensor, so it can be
decomposed in its longitudinal and transverse components by means of the projectors
$\mathcal{P}^{L}_{ij} = q_iq_j/q^2$ and $P^{T}_{ij} = \delta_{ij}-\mathcal{P}^{L}_{ij}$. 
It has been shown (see \cite{baym-1967} and \cite{ueda-book} for a pedagogical review),
that the static limit of transverse component of 
Eq. \eqref{correlator in fourier space after wick}
$\mathcal{C}_{ij}^{T}(\mathbf{q},\omega_n)\equiv (d-1)^{-1}\sum_{ij}\mathcal{P}_{ij}^{T}
(\mathbf{q},\omega_n)
\mathcal{C}_{ij}(\mathbf{q},\omega_n)/\hbar$ corresponds to the normal component of the system.
By replacing the replicated correlator given by Eqs. \eqref{correlazione gaussiana density-density}
and \eqref{correlazioni gaussiane fase-fase e fase-densita} into 
Eq. \eqref{correlator in fourier space after wick}, after burdensome (but standard)
algebra, we find
\begin{equation}
\begin{aligned}
n_n & = \lim_{q\rightarrow 0} 
\braket{\mathcal{C}^{T}(\mathbf{q},0)}_{\text{th+dis}} \\
&= \frac{1}{d\beta L^d}\sum_{\mathbf{k},\omega_m}\bigg[
\bigg(\frac{\hbar^2 k^2}{m}\bigg) \frac{E_k^2-\hbar^2\omega_m^2}{(\hbar^2\omega_m^2
	+E_q^2)^2} + n_0\bigg(\frac{\hbar^2 k^2}{m}\bigg)^2
\frac{E_k^2-3\hbar^2\omega_m^2}{(\hbar^2\omega_m^2+E_k^2)^3}
\gamma^2(\mathbf{k},\omega_m)
\bigg]\;.
\end{aligned}
\label{densita normale con disordine funzione di risposta}
\end{equation}
When the time scale of the disorder potential is frozen, i.e. for 
\begin{equation}
\gamma^2(\mathbf{q},\omega_n) = \gamma^2(\mathbf{q})\beta\delta_{n,0}\;, 
\label{disordine statico point-like}
\end{equation}
Eq. \eqref{densita normale con disordine funzione di risposta} becomes
\begin{equation}
n_n(n_0,T) = \frac{\beta}{d}\int \frac{d^d\mathbf{q}}{(2\pi)^d}
\frac{\hbar^2 q^2}{m} \frac{e^{\beta E_q}}{(e^{\beta E_q} -1)^2}
+\frac{4n_0}{d}\int \frac{d^d\mathbf{q}}{(2\pi)^d} \bigg(\frac{\hbar^2 q^2}{2m}\bigg)^2
\frac{\gamma^2(\mathbf{q})}{E_q^4}\;.
\label{componente normale per disordine statico formula generale}
\end{equation}
With the further simplification of a constant disorder correlator $\gamma^2$, the
equation above equates to Eq. \eqref{superfluid depletion}. So, 
Eq. \eqref{componente normale per disordine statico formula generale}
reduces corretly to the quenched disorder regime and agrees with results 
reported in \cite{tauber-1997,schakel-1997,giorgini-1994}.

In order to compute the condensate depletion through the replicated correlators in 
Eqs. \eqref{correlazione gaussiana density-density} and 
\eqref{correlazioni gaussiane fase-fase e fase-densita}, we rely upon the equation
\cite{fetter-book}
\begin{equation}
n = n_0 + \frac{1}{\beta\hbar}\int \frac{d^d\mathbf{q}}{(2\pi)^d} 
\sum_n\, e^{i\omega_n 0^+}\big[ \braket{\psi(\mathbf{r},\tau)\psi^*(\mathbf{r}',\tau')}
-n_0
\big]\;.
\label{depletion del condensato con funzione di green}
\end{equation}
Following the phase-density representation in Eq. \eqref{splitting del campo densita-fase},
we expand the fields $\psi$ and $\psi^*$ up to the linear order in the
fluctuations $\psi(\mathbf{r},\tau) \simeq \sqrt{n_0} [1 + \pi(\mathbf{r},\tau)/(2n_0) 
+ i\varphi(\mathbf{r},\tau)]$, such that
the resulting correlators in Eq. \eqref{depletion del condensato con funzione di green}
are Gaussian. In the Fourier space we have
\begin{equation}
\braket{\psi(\mathbf{q},\omega_n)\psi^*(\mathbf{q}',\omega_{n'})} -n_0 \simeq
\frac{1}{4n_0} \braket{\pi\pi'} -
\frac{i}{2}\braket{\pi\varphi'}
+\frac{i}{2}\braket{\varphi\pi'}
+n_0 \braket{\varphi\varphi'}
\label{correlatore campo-campo per la funzione di green}
\end{equation}
where the prime signal the dependence on $Q' = (\mathbf{q}',\tau')$
and, similarly to Eq. \eqref{densita di corrente definizione}, $\braket{\ldots}$
corresponds to a general average procedure. 
In order to
include the disorder contribution one simply has to consider the replicated
correlators in Eqs. \eqref{correlazione gaussiana density-density}
and \eqref{correlazioni gaussiane fase-fase e fase-densita}, where space-time
translational invariance restored by assuming 
Eq. \eqref{valori di aspettazione per il sistema pure}.

We then have
\begin{equation}
\begin{aligned}
n & =
n_0 + \int \frac{d^d\mathbf{q}}{(2\pi)^d}\bigg[
\frac{\epsilon_q+n_0\tilde{V}(q)}{2E_q}- \frac{1}{2} + \frac{\epsilon_q+n_0\tilde{V}
(q)}{E_q}\bigg(\frac{1}{e^{\beta E_q} -1}\bigg) \\
& \qquad \qquad \qquad \qquad + \frac{n_0}{\beta\hbar} \sum_m
\frac{\epsilon_q^2 - \hbar^2\omega_m}{(\hbar^2\omega_m^2 + E_q^2)^2}
\gamma^2(\mathbf{q},\omega_m)
\bigg]\;.
\end{aligned}
\label{depletion del condensato conto con la funzione di green}
\end{equation}
The first line represents the contribution coming from the pure system, reading 
Eq. \eqref{contributo gaussiano puro depletion temperatura nulla} at $T = 0 K$.
On the other hand, the second line accounts for the disorder contribution. For 
a point-like frozen disorder as in Eq. \eqref{disordine statico point-like},
one can verify that
\begin{equation}
n_{\gamma}(n_0) =
n_0 \int \frac{d^d\mathbf{q}}{(2\pi)^d} \bigg(\frac{\hbar^2q^2}{2m}\bigg)^2\,
\frac{\gamma^2(\mathbf{q})}{E_q^4}
\label{contributo di disordine generale densita totale} 
\end{equation}
By comparing the equation above and 
Eq. \eqref{componente normale per disordine statico formula generale}, at zero-temperature
the disorder contribution to the condensate depletion and the normal component of the system
are deeply related, namely
\begin{equation}
n_{\gamma}(n_0)= \frac{d}{4}n_n(n_0,T=0)
\end{equation}
%
%
%
%

\section{Conclusions and future perspectives}

In this review, we have reviewed the crucial aspects of a field-theory approach
to superfluid bosons moving in a random environment. Within the powerful framework 
of functional integration, we have recovered the important result on the superfluid and 
condensate depletion in presence of an external disorder 
\cite{huang-1992,giorgini-1994,schakel-1997}. 
Moreover, we have shown how the theoretical description can be 
generalized beyond the assumption of a static external 
disorder by using the replicated formalism \cite{navez-2007}. 
Obviously, our pedagogical overview has to left out other interesting
and crucial issues. For instance, we have pointed out in Sec. \ref{sec:3} that
our perturbative approach lacks self-consistency, since we are expanding above a 
uniform background also in presence of an external potential. 
In order to overcome this point, in \cite{khellil-2016,khellil-2016-2}
the authors present a self-consistent implementation relying upon the Hartee-Fock 
mean-field theory within the replica formalism. 
Moreover, our review does not include any discussion about the eventual occuring
of the superfluid-glass transition in three or lower dimensions. This is a crucial topic,
deserving a separate investigation.
Numerical simulations based on 
quantum Monte Carlo methods are likewise important, since they are \textit{ab-initio}
calculation and may shed light on unexpected effects. For instance, in 
\cite{giorgini-2002} the non-trivial relation between superfluidity and condensation is
numerically investigated, while in \cite{ng-2015} the worm algorithm provides
an estimation for critical exponents of the superfluid-glass transition in two-dimensions.

Another promising line of research concerns the possibility that disorder does not
always act in a \textit{parasitic} way, generating, on the contrary, surprising dynamical effects.
In the context of condensed matter theory, one of the most striking example is given
by the quantum Hall transition \cite{pruisken-1987}, where impurity scattering plays 
a crucial role. 
On the other hand, very recently
it was shown that a certain degree of random fluctuations, such as thermal noise, may 
enhance the transport properties of a particles ensemble. While Anderson localization
\cite{anderson-1959} acts to halt the flow of a certain quantity, 
in \cite{caruso-2015,potocnik-2018} the authors observe a boosting
of transport through optical fibers and superconducting circuits. The reason is the so-called
environment-assisted quantum trasport, which relies crucially upon the presence of
a certain degree of disorder. Indeed, while the coherence of the system is reduced by
propagating in a disordered medium, this also reduces the possibility of destructive interference
responsible for Anderson localization. 

This phenomenon can pave the way to novel interesting protocol to engineer more efficient
quantum devices, by tuning the coupling with the environment to enhance the 
trasport of a desired quantity. Recent experimental confirmations have been produced
by using a one-dimensional array of trapped ions \cite{maier-2019}. Thus, it would be
extremely interesting to review to effect in the context of ultracold atoms, both from
an experimental and theoretical point of view.

\vspace{6pt} 

\subsection*{Acknowlegdements}
LS acknowledges for partial support the FFABR grant of Italian
Ministry of Education, University and Research.
AC thanks Andrea Tononi for useful discussions and meaningful insight
on the manuscript.


\appendix
\section{ Diagonalization of a block circulant matrix}
\unskip
\label{app:A1}
It is possible to perform a block diagonalization of $\mathbb{M}^{-1}$ in
Eq. \eqref{propagatore gaussian circolante fourier}
by noticing its circulant structure. 
Indeed, $\mathbb{G}^{-1}$ 
is generated by cyclically permuting the elements (in this case $2\times 2$ matrices) of its
first row. In this sense, we can formally write down
\begin{equation}
\mathbb{M}^{-1}(Q) = \text{Circ}\big\lbrace
\tilde{\mathcal{M}}^{-1}(Q), \underbrace{\mathbb{B}(Q), \ldots ,\mathbb{B}(Q)}_{{}
	\text{$R-1$ times}}
\big\rbrace\;.
\label{propagatore replicato come circolante}
\end{equation}
It has been shown \cite{davis-book,olson-2014} that,
given $R$ the number of the generating matrices and $M$ their dimension,
every block circulant matrix $\mathbb{A}$ is block
diagonalized by the same unitary transformation. Indeed, one can verify that
\begin{equation}
(\mathbb{E}_R^{\dagger} \otimes 1_M) \mathbb{A} (\mathbb{E}_R \otimes 1_M)
= \text{diag}\lbrace \Lambda_1, \ldots , \Lambda_R\rbrace
\label{diagonalizzazione delle circolanti a blocchi}
\end{equation}
where the $\Lambda_{j}$ are the $2\times 2$ eigenblocks, specified by
$
\Lambda_{j} = \mathcal{M}^{-1} + \sum_{\alpha=1}^{R-1} \mathbb{B}\; w_R^{\alpha j}
$. The $\otimes$ symbol denotes the direct product of the two matrices.
In Eq. \eqref{diagonalizzazione delle circolanti a blocchi} the key ingredient is the
Fourier matrix, defined as
\begin{equation}
\mathbb{E}_R = \frac{1}{\sqrt{R}}
\begin{pmatrix}
1 		& 1 		& 1 		& \dots 	& \dots 	& 1 		\\
1 		& w_R		& w_R^2		& \dots		& \dots		& w_R^{R-1} \\
1	 	& w_R^2		& w_R^4		& \dots 	& \dots 	& w_R^{2(R-1)} \\
\vdots	& \vdots	& \vdots	& \ddots	& 			& \vdots \\
\vdots	& \vdots	& \vdots 	&			& \ddots	& \vdots \\
1 		& w_R^{R-1}	& w_R^{2(R-1)}	& \dots 	& \dots		& w_R^{(R-1)^2} \\
\end{pmatrix}
\label{matrice di fourier}
\end{equation}
with $w_R = \exp(2\pi i /R)$ the fundamental root of the unity.


\begin{thebibliography}{999}

\bibitem{cornell-1995} M. H. Anderson, J. R. Ensher, M. R. Matthews, C. E. Wieman
and E. A. Cornell, 
\textit{Observation of Bose-Einstein Condensation in a Dilute Atomic Vapor},
Science \textbf{269}, 5221 (1995).

\bibitem{ketterle-1995} K. B. Davis et \textit{al}.,
\textit{Bose-Einstein Condensation in a Gas of Sodium Atoms},
Phys. Rev. Lett. \textbf{75}, 3969 (1995).

\bibitem{pethick-book} C. J. Pethick and H. Smith, 
\textit{Bose-Einstein Condensation in Dilute Gases}, (Cambridge Univ. Press, 2011).

\bibitem{langen-2015} T. Langen, R. Geiger and J. Schmiedmayer, 
\textit{Ultracold atoms out of equilibrium},
Annu. Rev. Condens. Matt. Phys. \textbf{6}, 201-17 (2015).

\bibitem{fisher-1989} M. P. A. Fisher, P. B. Weichman, G. Grinstein
and D. S. Fisher, 
\textit{Boson localization and the superfluid-insulator transition},
Phys. Rev. B. \textbf{40}, 546 (1989).

\bibitem{greiner-2002} M. Greiner, O. Mandel, T. Esslinger, T. W. H\"ansch
and I. Bloch,
\textit{Quantum phase transition from a superfluid to a Mott insulator 
	in a gas of ultracold atoms}, Nature \textbf{415}, 39-44 (2002).

\bibitem{giamarchi-book} T. Giamarchi, \textit{Quantum Physics in One Dimensions},
(Clarendon Press, Oxford, 2003).

\bibitem{dalibard-review-2d} Z. Hadzibabic and J. Dalibard, 
\textit{Two-dimensional Bose fluids: an atomic physics perspective},
Rivista del Nuovo Cimento \textbf{34}, 389 (2011).

\bibitem{paredes-2004} B. Paredes, A. Widera, V. Murg, O. Mandel, S. Folling, 
I. Cirac, G. V. Shlyapnikov, T. W. H\"ansch, and I. Bloch, 
\textit{Tonks–Girardeau gas of ultracold atoms in an optical lattice},
Nature (London) \textbf{429}, 277 (2004).

\bibitem{kinoshita-2004} T. Kinoshita, T. Wenger, and D. S. Weiss, 
\textit{Observation of a One-Dimensional Tonks-Girardeau Gas}
Science \textbf{305}, 1125 (2004).

\bibitem{hadzibabic-2006} Z. Hadzibabic, P. Kr\"uger, M. Cheneau, 
B. Battelier and J. Dalibard,
\textit{Berezinskii-Kosterlitz- ouless crossover in a trapped atomic
gas}, Nature \textbf{441}, 1118 (2006).

\bibitem{cornell-2007} V. Schweikhard, S. Tung and E. A. Cornell,
\textit{Vortex proliferation in the Berezinskii-Kosterlitz- ouless regime on a two-dimensional
lattice of Bose-Einstein condensates}, Phys. Rev. Lett. \textbf{99}, 030401 (2007).

\bibitem{marquardt-2015} C. Neuenhahn and F. Marquardt,
\textit{Quantum simulation of expanding space–time with tunnel-coupled condensates},
New J. Phys. \textbf{17}, 125007 (2015).

\bibitem{brandt-2017} O. Fialko, B. Opanchuk, A. I. Sidorov, P. D. Drummond
and J. Brand,
\textit{The universe on a table top: engineering quantum decay of a relativistic 
	scalar field from a metastable vacuum}, J. Phys. B: At. Mol. Opt. Phys.
\textbf{50}, 024003 (2017).

\bibitem{braden-2018} J. Braden, M. C. Johnson, H. V. Peiris and S. Weinfurtner,
\textit{Towards the cold atoms analog of the false vacuum},
J. High En. Phys. 2018:14 (2018).

\bibitem{liberati-2006} S. Liberati,  M. Visser and S. Weinfurtner,
\textit{Analogue quantum gravity phenomenology from a two-component Bose-Einstein condensate},
Class. Quant. Grav. \textbf{23}, 3129 (2006).

\bibitem{sademelo-2018} D. M. Kurkcuoglu and C. S\'a de Melo,
\textit{Unconventional color superfluidity in ultra-cold 
	fermions: Quintuplet pairing, quintuple point and pentacriticality},
 	arXiv:1811.07272 (2018).

\bibitem{kamenev-review} A. Kamenev, \textit{Many-body theory of non-equilibrium systems},
in \textit{Nanophysics: Coherence and Transport}, p. 177-246 (Elsevier, Amsterdam, 2005).

\bibitem{ma-1985} M. Ma and P. A. Lee, \textit{Localized superconductors},
Phys. Rev. B \textbf{32}, 5658 (1985). 

\bibitem{ma-1986} M. Ma, B. I. Halpering and P. A. Lee,
\textit{Strongly disordered superfluids: Quantum fluctuations and critical behavior},
Phys. Rev. B \textbf{34}, 3136 (1986)

\bibitem{anderson-1959} P. W. Anderson, \textit{Absence of diffusion in certain random lattices},
Phys. Rev.  \textbf{109}, 1492 (1959).

\bibitem{damski-2003} B. Damski, J. Zakrzewski, L. Santos, P. Zoller and 
\textit{M. Lewenstein, Atomic Bose and Anderson Glasses in Optical Lattices}, 
Phys. Rev. Lett. \textit{91}, 080403 (2003).

\bibitem{schulte-2005} T. Schulte, S. Drenkelforth, J. Kruse, W. Ertmer, 
J. Arlt, K. Sacha, J. Zakrzewski 
and M. Lewenstein, 
\textit{Routes Towards Anderson-Like Localization of Bose-Einstein 
Condensates in Disordered Optical Lattices}, 
Phys. Rev. Lett. \textbf{95}, 170411 (2005).

\bibitem{billy-2008} J. Billy, V. Josse, Z. Zuo, A. Bernard, B. Hambrecht, P. Lugan,
D. Clement, L. Sanchez-Palencia, P. Bouyer and A. Aspect, 
\textit{Direct Observation of Anderson Localization of Matter-Waves
in a Controlled Disorder}, Nature \textbf{453}, 891 (2008).

\bibitem{roati-2008} G. Roati, C. D’Errico, L. Fallani, M. Fattori, C. Fort, 
M. Zaccanti, G. Modugno, M. Modugno and M. Inguscio, 
\textit{Anderson Localization of a Non-Interacting Bose-Einstein Condensate}, 
Nature \textbf{453}, 895 (2008).

\bibitem{dainty-1980} J. C. Dainty, 
\textit{An introduction to Gaussian speckle}, 
Proc. SPIE \textbf{243}, 2 (1980).

\bibitem{goodman-book} J. W. Goodman, 
\textit{Speckle Phenomena in Optics: Theory and Applications}, 
(Viva Books Private Limited, 2010).

\bibitem{lye-2005}J. E. Lye, L. Fallani, M. Modugno, 
D. S. Wiersma, C. Fort and M. Inguscio, \textit{A Bose-Einstein
condensate in a random potential}, Phys. Rev. Lett. \textbf{95}, 070401 (2005).

\bibitem{clement-2005} D. Cl\'ement, A. F. Var\'on, M. Hugbart, J. A. Retter, 
P. Bouyer, L. Sanchez-Palencia, D. M. Gangardt, G. V. Shlyapnikov and A. Aspect, 
\textit{Suppression of Transport of an Interacting Elongated Bose-Einstein Condensate in a Random 
Potential}, Phys. Rev. Lett. \textbf{95}, 170409 (2005).

\bibitem{ghabour-2014} M. Ghabour and A. Pelster,
\textit{Bogoliubov theory of dipolar Bose gas in a weak random potential},
Phys. Rev. A \textit{90}, 063636 (2014).

\bibitem{stoof-book} H. T. C. Stoof, D. B. M. Dickerscheid and K. Gubbels, 
\textit{Ultracold Quantum Fields} (Springer, Dordrecht, 2009).

\bibitem{salasnich-2016} L. Salasnich and F. Toigo, 
\textit{Zero-point energy of ultracold atoms},
Phys. Rep. \textbf{640}, 1-29 (2016).

\bibitem{huang-1992} K. Huang and H.-F. Meng,
\textit{Hard-Sphere Bose Gas in Random External Potential},
Phys. Rev. Lett. \textbf{69}, 644 (1992).

\bibitem{giorgini-1994} S. Giorgini, L. Pitaevskii and S. Stringari,
\textit{Effects of disorder in a dilute Bose gas}
Phys. Rev. B \textbf{49}, 18 (1994).

\bibitem{tauber-1997} U. Ta\"uber and D. R. Nelson, 
\textit{Superfluid bosons and flux liquids: disorder, thermal fluctuations,
	and finite-size effects}, Phys. Rep. \textbf{289},
157 (1997).

\bibitem{lopatin-2002} A. V. Lopatin and V. M. Vinokur,
\textit{Thermodynamics of the Superfluid Dilute Bose Gas with Disorder},
Phys. Rev. Lett. \textbf{88}, 235503 (2002).

\bibitem{falco-2007} G. M. Falco, A. Pelster and R. Graham, 
\textit{Thermodynamics of a Bose-Einstein condensate with weak disorder}
Phys. Rev. A \textbf{75}, 063619
(2007).

\bibitem{giamarchi-1988} T. Giamarchi and H. J. Schulz, 
\textit{Anderson localization and interactions in one-dimensional metals},
Phys. Rev. B \textbf{37}, 325 (1988).

\bibitem{navez-2007} P. Navez, A. Pelster and R. Graham, 
\textit{Bose condensed gas in strong disorder potential with arbitrary correlation length},
App. Phys. B \textbf{86}, 395-398 (2007).

\bibitem{yukalov-2007} V. I. Yukalov andd R. Graham,
\textit{Bose-Einstein condensed systems in random potentials},
Phys. Rev. A \textbf{75}, 023619 (2007).

\bibitem{falco-2009} G. M. Falco, T. Nattermann and V. L. Pokrovsky,
\textit{Weakly interacting Bose gas in a random environment},
Phys. Rev. B \textbf{80}, 104515 (2009).

\bibitem{khellil-2016} T. Khellil, A. Balaz and A. Pelster,
\textit{Analytical and numerical study of dirty bosons in a quasi-one-dimensional harmonic trap},
New J. Phys. \textbf{18}, 063003 (2016).

\bibitem{khellil-2016-2} T. Khellil and A. Pelster, 
\textit{Hartree-Fock Mean-Field Theory for Trapped Dirty Bosons},
J. Stat. Mech. 063301 (2016).

%
%
\bibitem{altland-book} A. Altland and B. Simons, \textit{Condensed Matter Field Theory}
(Cambridge University Press, 2010).

\bibitem{hertz-1985} J. Hertz, \textit{Disordered Systems}, Phys. Scr. \textbf{1985}, 1 (1985).

\bibitem{nelson-1990} D. R. Nelson and P. Le Doussal, 
\textit{Correlations in flux liquids with weak disorder}, Phys. Rev. B \textbf{42}, 16 (1990).

\bibitem{schakel-1997} A. M. J. Schakel, 
\textit{Quantum critical behavior of disordered superfluids}, 
Phys. Lett. A \textbf{224}, 287 (1997).

\bibitem{lubensky-1975} T. C. Lubensky, 
\textit{Critical properties of the random-spin model from the $\epsilon$-expansion},
Phys. Rev. B \textbf{11}, 9 (1975).

\bibitem{grinstein-1976} G. Grinstein and A. Luther, 
\textit{Applications of the renormalization group to phase transition in disordered
systems},
Phys. Rev. B \textbf{13}, 3 (1976).

\bibitem{schakel-book} A. M. J. Schakel, \textit{Boulevard of Broken Symmetries:
	Effective Field Theories of Condensed Matter}, (World Scientific, Singapore, 2008).


\bibitem{boronat-2009} G. E. Astrakharchik, J. Boronat, J. Casulleras, 
I. L. Kurbakov, and Yu. E. Lozovik, 
\textit{Equation of state of a weakly interacting two-dimensional Bose 
	gas studied at zero temperature by means of quantum Monte Carlo methods},
Phys. Rev. A \textbf{79}, 051602 (2009).

\bibitem{tononi-2018} A. Tononi, A. Cappellaro and L. Salasnich,
\textit{Condensation and superfluidity of dilute Bose gases with finite-range interaction},
New J. Phys. \textbf{20}, 125007 (2018).

\bibitem{salasnich-2017} L. Salasnich,
\textit{Nonuniversal Equation of State of the Two-Dimensional Bose Gas},
Phys. Rev. Lett. \textbf{118}, 130402 (2017).

\bibitem{tononi-2019} A. Tononi, 
\textit{Zero-temperature equation of state of a two-dimensional 
bosonic quantum fluid with finite-range interaction},
Condens. Matter \textbf{4(1)}, 20 (2019). 

\bibitem{lewenstein-2006} J. Wehr, A. Niederberger, L. Sanchez-Palencia and
M. Lewenstein, 
\textit{Disorder versus the Mermin-Wagner-Hohenberg effect: 
From classical spin systems to ultracold atomic gases}, Phys. Rev. B \textbf{74},
224448 (2006).

\bibitem{boudjema-2013} A. Boudjemaa and G.V. Shlyapnikov,
\textit{Two-dimensional dipolar Bose gas with the roton-maxon excitation spectrum},
Phys. Rev. A \textbf{87}, 025601 (2013).

\bibitem{boudjema-2015} A. Boudjemaa, 
\textit{Two-dimensional dipolar bosons with weak disorder},
Phys. Lett. A \textbf{379}, 2484-2487 (2015).

\bibitem{landau-book} L. D. Landau and E. M. Lifshitz,
\textit{Statistical Physics 2}, (Pergamon Press, Oxford, 1987).

\bibitem{khalatnikov-book} I. M. Khalatnikov, 
\textit{An Introduction to the Theory of Superfluidity}
(Westwiew Press, Oxford, 2000).

\bibitem{fisher-1973} M. E. Fisher, M. N. Barber and D. Jasnow,
\textit{Helicity, Modulus, Superfluidity and Scaling in Isotropic Systems},
Phys. Rev. A \textbf{8}, 1111 (1973).

\bibitem{taylor-2006} E. Taylor, A. Griffin, N. Fukushima and Y. Ohashi,
\textit{Pairing fluctuations and the superfluid density through the BCS-BEC crossover},
Phys. Rev. A \textbf{74}, 063626 (2006)

\bibitem{anderson-1975} S. F. Edwards and P. W. Anderson,
\textit{Theory of spin glasses}, J. Phys. F \textbf{5}, 965-974 (1975).

\bibitem{parisi-book} M. Mezard, G. Parisi and M. Virasoro, 
\textit{Spin Glass Theory and Beyond: An Introduction to the Replica Method and Its Applications}
(World Scientific, 1987).

\bibitem{parisi-review} G. Parisi, \textit{Glasses, replicas and all that}, in
Les Houches - Ecole d' \'et\'e de Physique Th\'eorique, Vol. \textit{77} (Elsevier, 2004).

\bibitem{mermin-1966} N. D. Mermin and H. Wagner, 
\textit{Absence of Ferromagnetism or antiferromagnetism in one- or two- dimensional 
	isotropic heisenberg models}, Phys. Rev. Lett. \textbf{17}, 1133 (1966)

\bibitem{hohenberg-1967} P. C. Hohenberg,
\textit{Existence of Long-Range Order in One and Two Dimensions},
Phys. Rev. \textbf{158}, 383 (1967).

\bibitem{villain-1975} J. Villain,
\textit{Theory of one- and two-dimensional magnets with an easy magnetization plane. 
	II. The planar, classical, two-dimensional magnet},
Journal de Physique \textbf{36}, 581 (1975).

\bibitem{kadanoff-1977} J. V. Jos\'e, L. P. Kadanoff,
S. Kirkpatrick and D. R. Nelson, 
\textit{Renormalization, vortices and symmetry-breaking perturbations in the two-dimensional 
	planar model},
Phys. Rev. B \textbf{16}, 1217 (1977).

\bibitem{andersen-2004} J. O. Andersen, 
\textit{Theory of the weakly interacting Bose gas},
Rev. Mod. Phys. \textbf{76}, 599 (2004).

\bibitem{fetter-book} A. L. Fetter and J. D. Walecka,
\textit{Quantum Theory of Many-particle systems}
(Dover, 2003).

\bibitem{baym-1967} G. Baym,
\textit{Microscopic Description of Superfluidity},
in \textit{Mathematical Methods in Solid State and Superfluid Theory}
(Springer, 1967).

\bibitem{ueda-book} M. Ueda, 
\textit{Fundamentals and New Frontiers in Bose-Einstein Condensations},
(World Scientific, 2010).

\bibitem{davis-book} P. J. Davis, \textit{Circulant Matrices} (Wiley, 2$^{\text{nd}}$ ed., 
New York, 1979).

\bibitem{olson-2014} B. Olson, S. Shaw, C. Shi, C. Pierre and R. Parker,
\textit{Circulant Matrices and Their Application to Vibration
Analysis}
Applied Mechanics Review \textbf{66}, 4 (2014). 

\bibitem{giorgini-2002} G. E. Astrakharchik, J. Boronat, J. Casulleras and S. Giorgini,
\textit{Superfluidity versus Bose-Einstein condensation in a Bose gas with disorder},
Phys. Rev. A \textbf{66}, 023603 (2002).

\bibitem{ng-2015} R. Ng and E. S. Sorensen, 
\textit{Quantum Critical Scaling of Dirty Bosons in Two Dimensions},
Phys. Rev. Lett. \textbf{114}, 255701 (2015).

\bibitem{pruisken-1987} A. M. M. Pruisken, \textit{Field theory, scaling and the
localization problem} in \textit{The Quantum Hall Effect} (Springer-Verlag, 1987).

\bibitem{caruso-2015} S. Viciani, M. Lima, M. Bellini and F. Caruso,
\textit{Observation of noise-assisted transport in an all-optical cavity-based
network}, Phys. Rev. Lett. \textbf{115}, 083601 (2015).

\bibitem{potocnik-2018} A. Potocnik \textit{et al.}, 
\textit{Studying light-harvesting models with superconducting circuits},
Nature Commun. \textbf{9}, 904 (2018).

\bibitem{maier-2019} C. Maier \textit{et al.},
\textit{Environment-Assisted Quantum Transport in a 10-qubit Network},
Phys. Rev. Lett. \textbf{122}, 050501 (2019).

\end{thebibliography}
\end{document}